\documentclass[final,12pt]{elsarticle}
\usepackage{subcaption}
\usepackage{amsmath}
\usepackage{amsbsy} 
\usepackage{tabularx}
\usepackage{stmaryrd}
\usepackage{enumitem}
\usepackage{siunitx}
\usepackage[numbers]{natbib}

\usepackage{amssymb}
\usepackage{xcolor}
\usepackage{geometry}
\makeatletter
\def\ps@pprintTitle{%
  \let\@oddhead\@empty
  \let\@evenhead\@empty
  \def\@oddfoot{} 
  \def\@evenfoot{} 
}
\makeatother

\begin{document}

\begin{frontmatter}



\title{Uncertainty Quantification in Multiscale Modeling of Polymer Composite Materials Using Physically Recurrent Neural Networks}

\author[a,b]{N. Kovács}
\author[b]{I.B.C.M. Rocha}
\author[b]{F.P. van der Meer}
\author[a]{C. Furtado}
\author[a]{P.P. Camanho}

\address[a]{{INEGI, Faculdade de Engenharia, Universidade do Porto}, {Rua Dr. Roberto Frias, 400, 4200-465 Porto}, {Portugal}}
\address[b]{{Delft University of Technology, Department of Civil Engineering and Geosciences}, {P.O.Box 5048}, 2600GA, Delft, {The Netherlands}}
\begin{abstract}
This study investigates whether Physically Recurrent Neural Networks (PRNNs), a recent surrogate model for heterogeneous materials, trained on a micromodel with fixed material parameters, can maintain accuracy for varying material properties without retraining, and propagate uncertainty in a multiscale framework. Unlike conventional RNNs, where parameter changes require training or explicit inclusion of material properties as extra input features, PRNNs embeds material models in their material layer that allow for modification of material parameters after training. When adjusting material properties dynamically according to the input during testing, PRNN shows high accuracy across a wide range of parameters. Therefore the surrogate can be applied to multiscale uncertainty quantification (UQ). Compared to the full-order simulations on an overly coarse mesh, the PRNN-driven model reduces simulation time by over 7000 times while accurately capturing highly nonlinear evolution of the probability density for the macroscopic response as a result of a given distribution for microscale material parameters. A PRNN-driven UQ is demonstrated on a more accurate finer mesh that would be computationally infeasible with the full-order model.
\end{abstract}




\end{frontmatter}


\section{Introduction}
\label{}
The efficient use of polymer composite materials in structural applications relies on the capability of simulating their complex, non-linear behavior across different length scales. However, it is challenging to model the response of composite materials due to their heterogeneity. The two-scale Finite Element approach (FE$^2$), a concurrent multiscale modeling method, is an accurate way to capture the nonlinear behavior of composite materials \cite{Schröder2014,FEYEL2000309,FEYEL20033233}. Although FE$^2$ bypasses the need to explicitly compute the constitutive relation at the macroscale, the computational homogenization from microscale is very computationally demanding. To make matters worse, performing these multiscale simulations only once for a single microstructural configuration is often not enough.

To achieve optimal structural performance, composites can be designed by tailoring microstructural parameters. In order to optimize composite materials, a good understanding of the effect of variations in geometric configurations and constituent properties on the overall structural response is required. Modeling parameter variations is not only important in an optimization framework, but also when accounting for uncertainties in the structural response, which can originate from manufacturing errors such as voids, fiber misalignment or fiber waviness \cite{xinbao,HUANG20051964,CHUN2001125}.

Either in optimization or uncertainty quantification frameworks, an extensive parameter study such as Monte Carlo sampling is infeasible due to the already excessive computational time of even a single multiscale simulation. Alternatively, other stochastic methods can be applied. Perturbation-based stochastic finite element methods have been shown to perform well for a small interval of parameter uncertainty, while spectral stochastic finite element methods can be applied for larger variations or spatial uncertainties \cite{ZHOU2022115132}. However, these methods could become computationally expensive when the number of random variables increases. As an alternative, surrogate models can be used to assess the reliability of homogenized elastic properties when combinations of material and geometrical uncertainties are considered \cite{OMAIREY2018204,OMAIREY2019106925}. Although these approaches successfully reduce computational time, they are generally designed for specific parameter ranges and can face challenges when extrapolating beyond this space. Moreover, many of these methods primarily focus on linear elastic properties. Some nonlinear applications have been explored, such as in \cite{clement2012computational,CLEMENT201361}, where stochastic homogenization of hyperelastic microstructures with random geometry were demonstrated, using a non-concurrent approach and complex offline computations for generating the studied properties.

Considering deterministic properties, neural networks have been successfully used as surrogate models in accelerating multiscale modeling. At the microscale, to overcome the computational bottleneck of FE$^2$, the full-order micromodel is replaced by a neural network that predicts homogenized response at the integration points. To capture path-dependent behavior, Recurrent Neural Networks (RNNs) have been used to store history-related parameters over time \cite{GHAVAMIAN2019112594,WU2020113234}. A major drawback of RNNs is the excessive amount of training data they require to cover every possible loading scenario studied. To improve the extrapolation capabilities of RNNs, introducing physical laws in various ways into the network have been explored. In recent works, physics-augmented networks have been introduced, where the networks are constructed for a specific material model automatically fulfilling physical constraints \cite{SCHOMMARTZ2025117592,klein2023parametrisedpolyconvexhyperelasticityphysicsaugmented,KLEIN2022115501}. Although a reduction in training effort and increased generalization capabilities are achieved, they are material model specific. Moreover, for more advanced constitutive laws, they could become significantly more complex. In \cite{LIU20191138}, Deep Material Networks (DMNs) that are built up from physics based building blocks were introduced, and nonlinear responses of microstructures were accurately captured when trained on linear data.

Another physics-based approach using Physically Recurrent Neural Networks (PRNNs) integrate the same constitutive models from the full-order micromodel into their hidden layer. The PRNN introduced in \cite{MAIA2023115934} showed excellent accuracy in both micro- and multiscale settings when predicting the response from a microstructure with elastic inclusions in an elastoplastic matrix. A significant reduction in computational time compared to using FE$^2$ was achieved while requiring a smaller training dataset compared to state-of-the-art RNNs \cite{MAIA2023115934}. The network was later generalized to a 3D finite strain framework, considering rate-dependent materials \cite{MAMAIA2024105145}. The effect of microscale damage was also incorporated in \cite{kovács2024physicallyrecurrentneuralnetworks}. Most recently, PRNNs have been applied to accurately simulate off-axis experiments on thermoplastic composites \cite{maia2025surrogatebasedmultiscaleanalysisexperiments}. 

While using RNNs as surrogates for deterministic values is widely studied, extending these to uncertainty quantification frameworks is not. In \cite{TANG2022115726}, a data-driven method using wide and deep neural networks was introduced to link microstructural variations to the macro-scale tensile response in woven composites. Although trained on physics-based data, its capabilities to extrapolate outside the training range could be limited if physical constraints aren’t explicitly included. The DMNs were used in \cite{HUANG2022115197}, where a microstructure-guided deep material network was developed. Using a small amount of base DMNs, microstructures with different material and geometrical parameters were accurately predicted. The model was retrained to include different parameter sets in the training data along with parameter-related input features, and focuses on interpolation within the trained space.

Compared to conventional NN-based surrogates where parameters of the networks remain fixed after training, PRNNs offer a potential advantage for uncertainty propagation, because the direct implementation of the constitutive models allows for material parameter modifications after training. This motivates the study presented here, which explores whether the PRNN in \cite{MAIA2023115934} trained on a fixed RVE can generalize to unseen microscale material properties without retraining. For this, the multiscale model considered in this work is introduced and the training process of the PRNN is presented first. After training the selected PRNN on fixed parameters, its performance at the microscale is evaluated for varying material properties. Finally, the network's accuracy in the multiscale setting is demonstrated through uncertainty quantification using Monte Carlo simulations where the probabilistic macroscopic response as a result of uncertainty on microscopic material parameters is evaluated. 

\section{PRNN as a surrogate for microstructures with fixed parameters}
\label{}


This section presents a brief introduction to the computational challenge associated with FE$^2$ simulations. Afterwards, the full-order micromodel with fixed parameters used to generate the datasets is discussed. The PRNN is then briefly presented as a surrogate model replacing the full-order micromodel in the multiscale simulations. Finally, the training procedure for the PRNN is described, which is used in the rest of the work for evaluating the network's ability to extrapolate to other microscale material properties.

\subsection{The FE$^\text{2}$ method}

In the FE$^2$ method, finite element simulations are coupled across two scales by nesting a FE micromodel into each integration point of a macroscale FEM. The macrostructure is treated as a homogeneous structure, while the micromodel captures the material heterogeneity. This micromodel is assumed to be a representative volume element (RVE), which is evaluated as a boundary value problem (BVP) using microscale constitutive models for each constituent. The resulting microscale stress field is homogenized into a macroscale stress response, which is then used to compute the global internal forces at the macroscale. Using a nonlinear iterative solver, the BVP at the microscale has to be solved at each integration point, at each iteration, and for every time step, making the FE$^2$ method computationally very expensive. Moreover, in the case of an optimization framework or uncertainty quantification when the microscale parameters are modified, the entire FE$^2$ simulation has to be re-computed for each set of parameters, further increasing the already excessive computational cost.



\newpage
\subsection{Micromodel used for generating training dataset}

\begin{figure}[!h]
\centering
\includegraphics[scale=0.1]{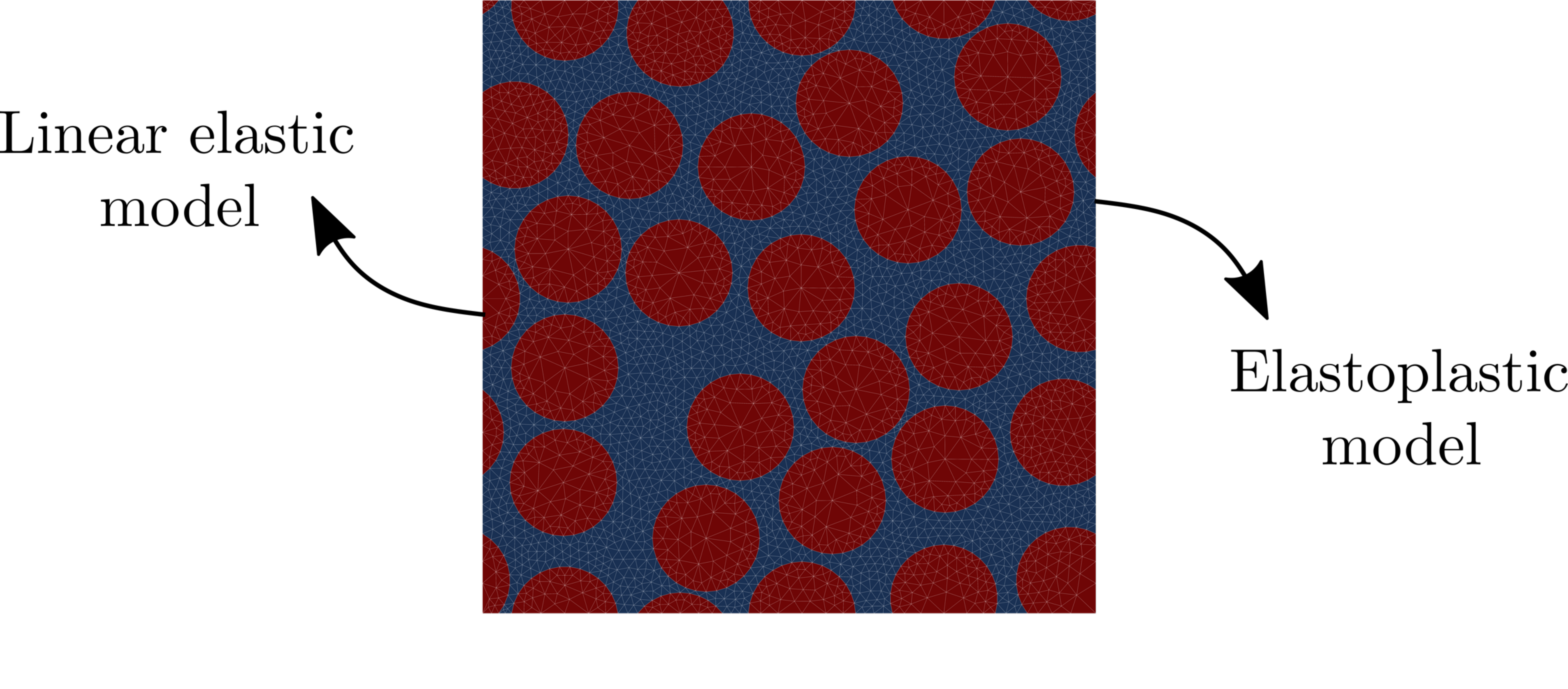}
\caption{\centering Finite element model of the full-order micromodel}
\label{fig:micro}
\end{figure}

In this paper, the microstructure shown in Figure \ref{fig:micro} is considered in the full-order simulations. It consists of 25 periodically arranged fibers with diameter of $\SI{5}{\micro\metre}$, modeled as linear elastic, embedded in a matrix, which is modeled using von Mises yield criterion. A fiber volume fraction of 0.6 is used. 

The material properties reported in \cite{MAIA2023115934} are used as a benchmark to obtain the training data, which are: the Young's modulus of the matrix and the fiber ($E_m = 3 130$ MPa and $E_f = 74 000$ MPa), Poisson's ratio of the matrix and the fiber ($\nu_m = 0.37$ and $\nu_f = 0.2$), and the yield strength of the matrix ($\sigma_y = 64.8$ MPa) with yield function:
\begin{equation}\label{eq:y}
   \sigma_y = 64.8 - 33.6 e^{-\varepsilon_{eq}^p/0.0003407}
\end{equation}
with equivalent plastic strain $\varepsilon_{eq}^p$ given by:
\begin{equation}
   \varepsilon_{eq}^p=\sqrt{\frac{2}{3} \pmb{\varepsilon}^p:\pmb{\varepsilon}^p}
\end{equation}
with $\pmb{\varepsilon}^p$ being the plastic strain. The training dataset is generated by subjecting the micromodel with these properties to non-proportional, non-monotonic loading scenarios, ensuring that the PRNN is exposed to a diverse loading space. For these load paths, strains are sampled from Gaussian Processes (GPs), again following \cite{MAIA2023115934}. The role of the PRNN is to replace solving this micromodel at every integration point of the macrostructure, which is responsible for the computational bottleneck that limits the widespread use of the FE$^2$ method in industrial applications.

\subsection{PRNN architecture}

The key idea behind PRNNs is the direct implementation of the constitutive models from the RVE into the hidden layer of the network. This approach introduces physics-based recurrency, as internal variables are stored within the network that capture the response of path-dependent materials. As demonstrated in \cite{MAIA2023115934,kovács2024physicallyrecurrentneuralnetworks}, the PRNN has to include all nonlinear models in the RVE it replaces for accurate predictions. Since the micromodel used in this work contains only one nonlinear model, namely the elastoplastic model for the matrix, the hidden layer of the network consists of fictitious material points with the same embedded constitutive law.

\begin{figure}[!ht]
\centering
\includegraphics[scale=0.8]{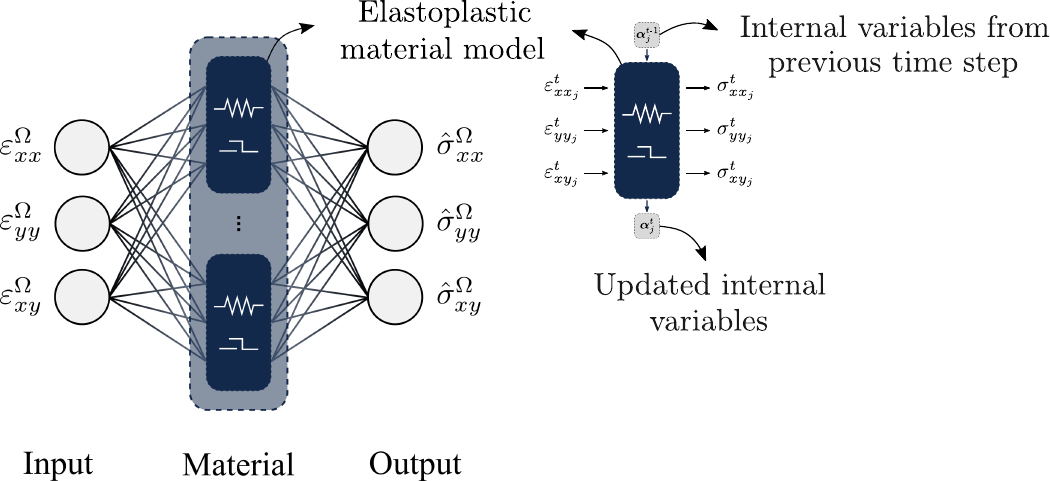}
\caption{\centering Physically Recurrent Neural Network with elastoplastic material model}
\label{fig:prnn}
\end{figure}

The architecture of the network is shown in Figure \ref{fig:prnn}. In its input, the network receives the macroscale strain. A linear encoder maps these strain values into sets of micro strains that are input to the fictitious points in the material layer. Figure \ref{fig:prnn} shows a bulk point with the operation of computing the local stresses using the elastoplastic material model used in the full-order micromodel, while also storing internal state parameters to keep track of history. A decoder then transforms these local stresses to the macroscale homogenized stress prediction of the network, which is then used in the global equilibrium equation. For more details on the network's implementation, the reader is referred to \cite{MAIA2023115934}.

\subsection{Model selection}

\begin{figure}[!h]
  \centering
  \begin{subfigure}[b]{0.45\textwidth}
    \centering
    \includegraphics[width=\textwidth]{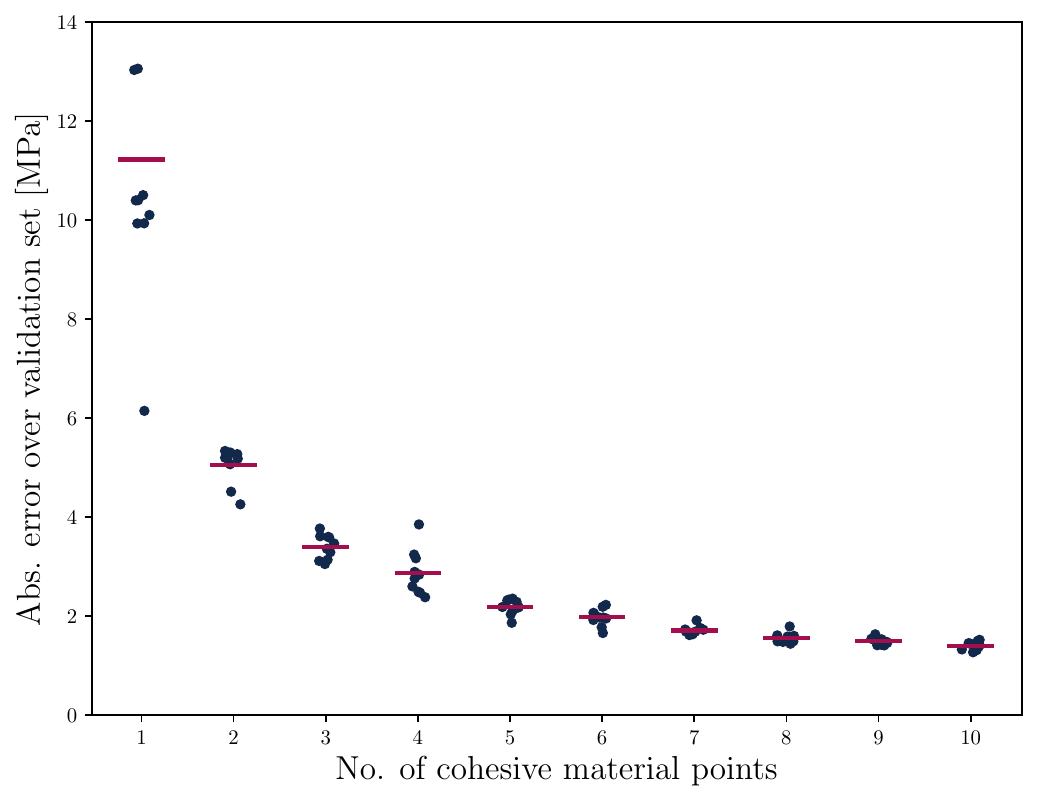}
    \caption{\centering PRNN trained on 128 GP curves}
    \label{fig:0ms}
  \end{subfigure}
  \hfill
  \begin{subfigure}[b]{0.45\textwidth}
    \centering
    \includegraphics[width=\textwidth]{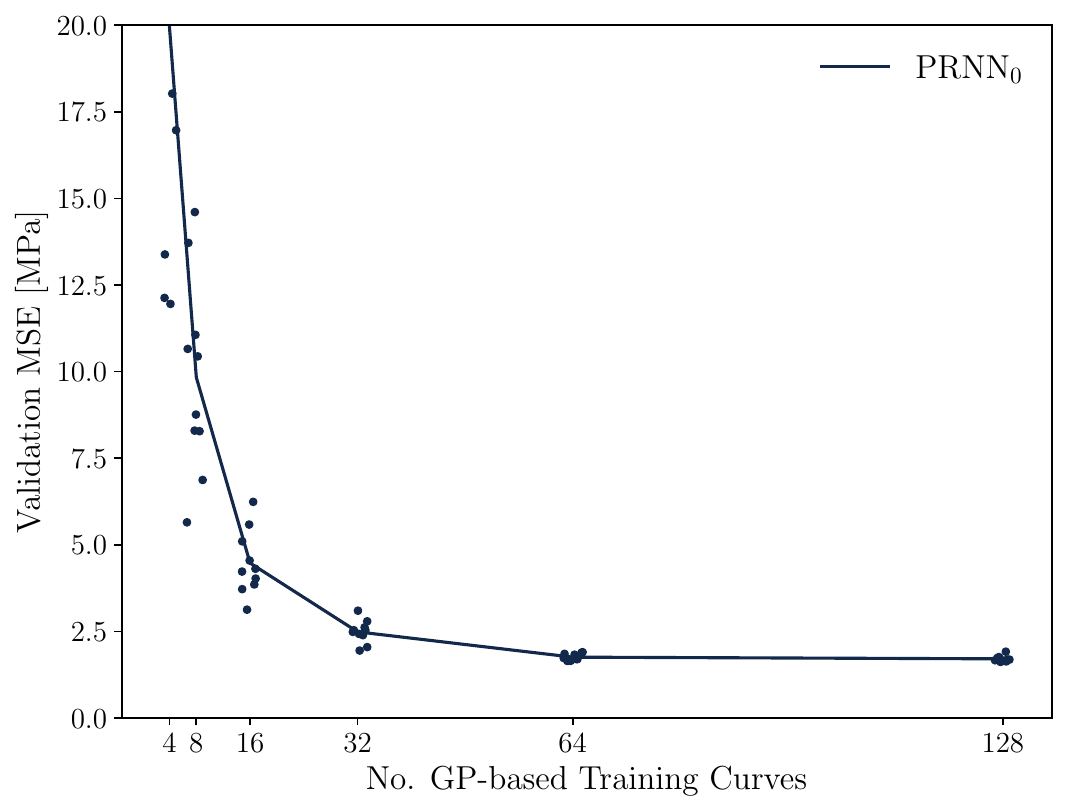}
    \caption{\centering PRNN with 7 bulk points}
    \label{fig:0lc}
  \end{subfigure}
\caption{\centering Validation error for PRNN trained on a fixed set of material properties}
  \label{fig:p0}
\end{figure}

During training, the material properties ($E_m$, $\nu_m$, $\sigma_y$) in the material layer of the PRNN are set to the reference values that are used in the micromodel that generates the training data. Model selection for this PRNN trained on a fixed set of material properties is performed by analyzing the performance of the network across different training set and material layer sizes. To determine the optimal size of the material layer, networks are trained on 128 GP-based curves with the number of points in the material layer varying from 1 to 10. Figure \ref{fig:0ms} shows the mean squared error (MSE) on a validation set comprising 56 GP-based curves for 10 different initializations, with the purple line representing the mean value at each material layer size. The network with 7 points is selected, as no significant reduction in mean validation MSE or scatter in errors can be observed with larger networks. Using the selected material layer size, the PRNN was trained on different dataset sizes, ranging from 4 to 128 GP-based curves. Figure \ref{fig:0lc} displays the MSE values on the validation set across the various training data sizes, with the solid lines representing the mean MSE values for each PRNN. A training set size of 64 paths was selected for the PRNN.

\begin{figure}[!h]
  \centering
  \begin{subfigure}[b]{0.45\textwidth}
    \centering 
    \includegraphics[width=\textwidth]{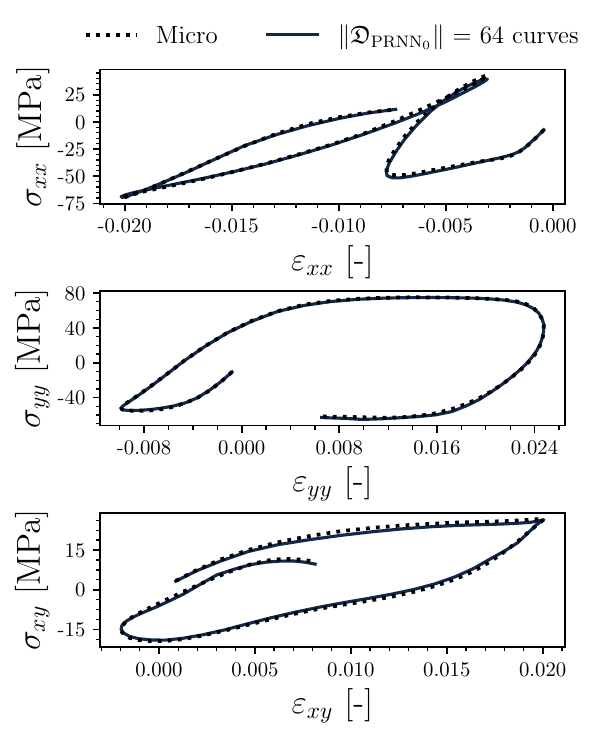}
    \caption{\centering Prediction on a GP-based curve}
    \label{fig:gp}
  \end{subfigure}
  \hfill
  \begin{subfigure}[b]{0.45\textwidth}
    \centering
    \includegraphics[width=\textwidth]{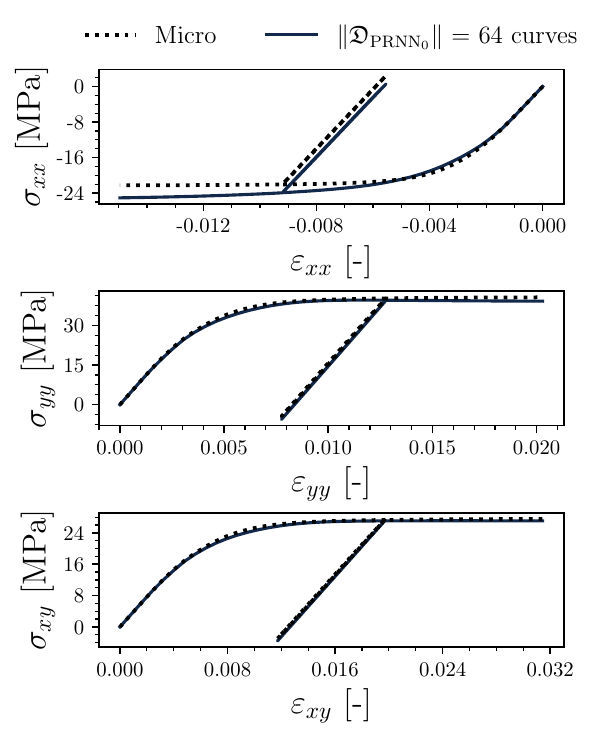}
    \caption{\centering Prediction on a non-monotonic curve}
    \label{fig:nm}
  \end{subfigure}
\caption{\centering Prediction of PRNN on representative test curves with fixed properties}
  \label{fig:oldie}
\end{figure}

The network's prediction on a curve from a test dataset containing the same type of curves as those used for training (GP-based) is shown in Figure \ref{fig:gp}. The high accuracy observed on this prediction also holds when predicting non-monotonic responses when loading in a fixed direction with one cycle of unloading, as seen in Figure \ref{fig:nm}. As established in \cite{MAIA2023115934}, the PRNN can accurately extrapolate to different loading scenarios when the microstructure properties remain the same as those used for training. We now extend the study to assess whether the accuracy of the PRNN trained on a fixed set of parameters is maintained when predicting responses from microstructures with varying parameters by also varying these parameters in the constitutive model in the material layer but without retraining the network (i.e. encoder and decoder weights are not modified).

\section{Extrapolating from a pretrained PRNN}
\label{}

\begin{figure}[htbp]
    \centering
    \begin{minipage}{\textwidth}
    \centering
    \begin{subfigure}{0.32\textwidth}
        \centering
        \includegraphics[width=\linewidth]{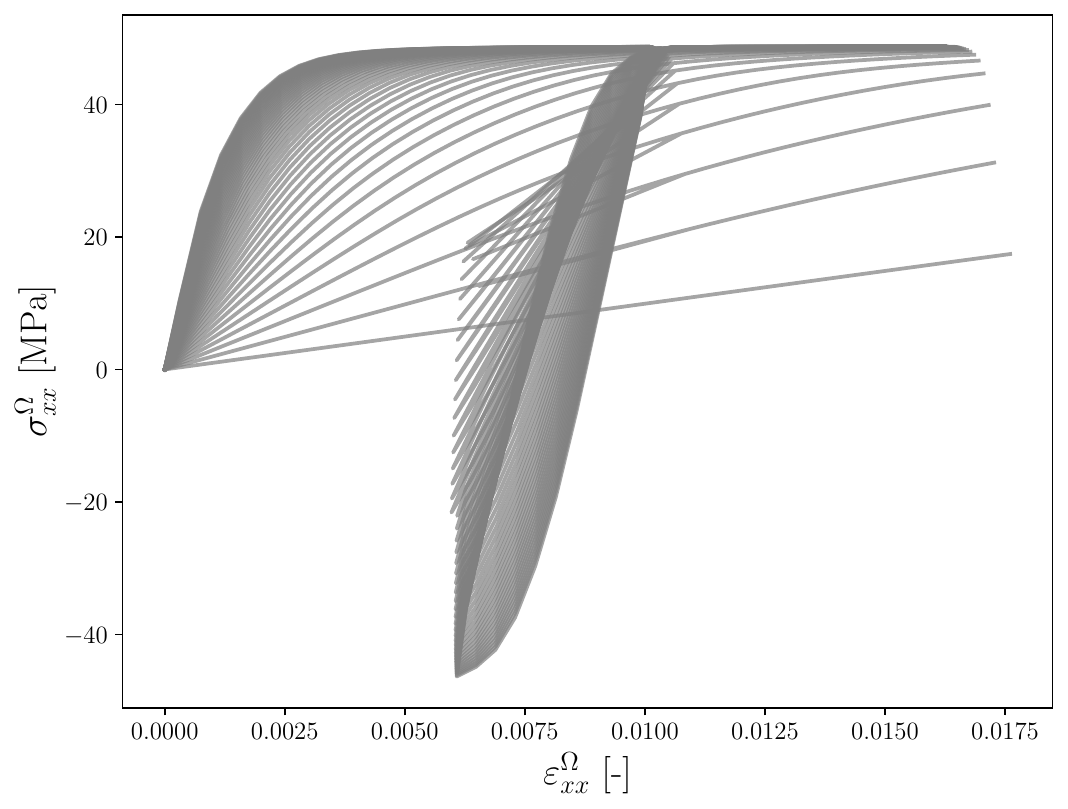}
        \caption{Young's modulus of matrix}
        \label{em}
    \end{subfigure}
    \hfill
    \begin{subfigure}{0.32\textwidth}
        \centering
        \includegraphics[width=\linewidth]{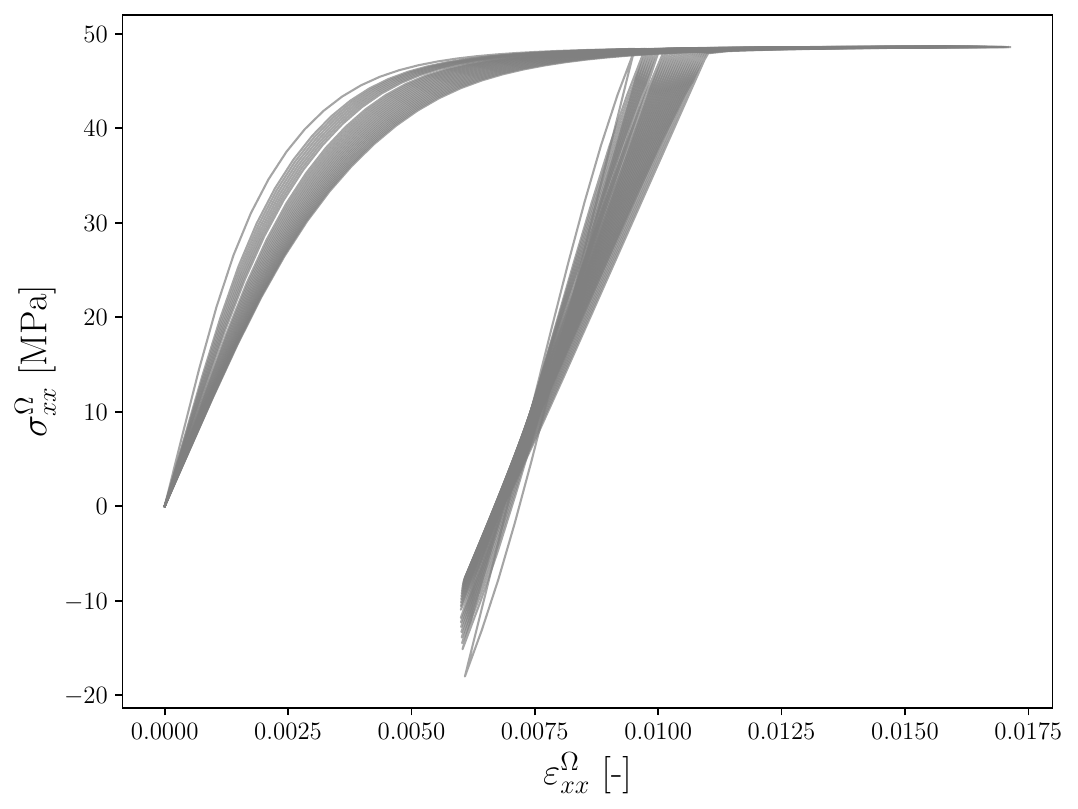}
        \caption{Poisson's ratio of matrix}
        \label{nm}
    \end{subfigure}
    \hfill
    \begin{subfigure}{0.32\textwidth}
        \centering
        \includegraphics[width=\linewidth]{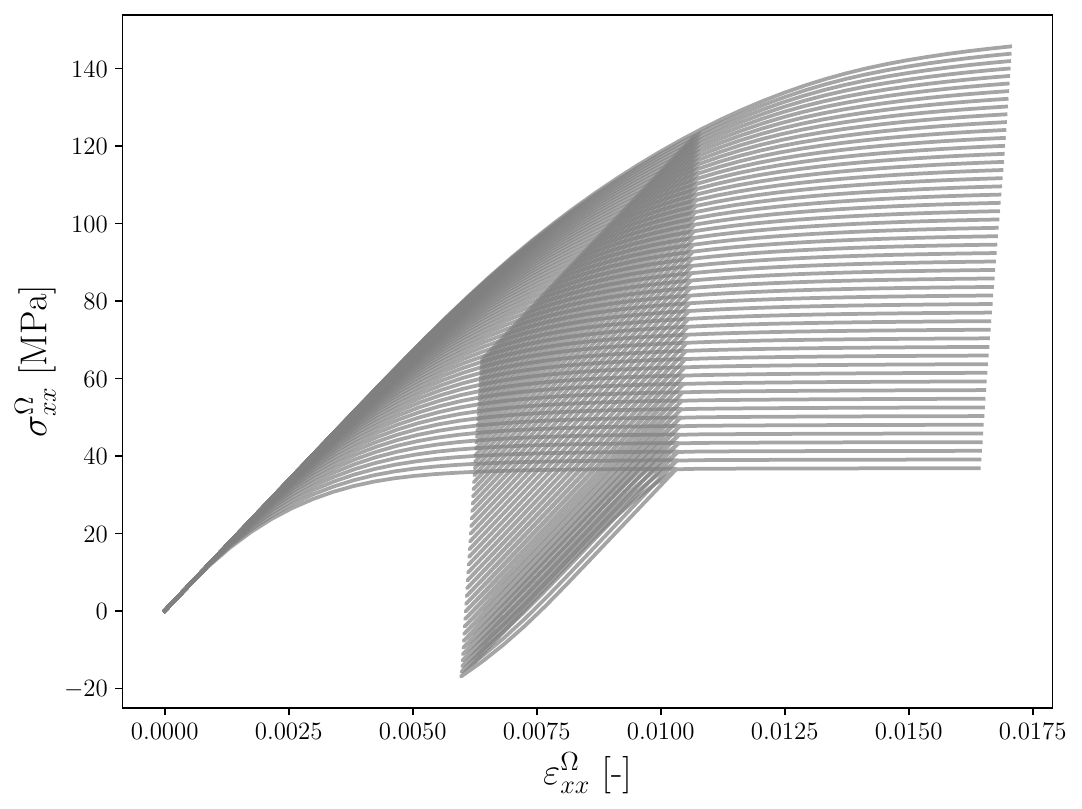}
        \caption{Yield strength of matrix}
        \label{plc}
    \end{subfigure}
    \end{minipage}
    \vskip\baselineskip
    \begin{minipage}{0.66\textwidth}
    \centering
    \begin{subfigure}{0.49\textwidth}
        \centering
        \includegraphics[width=\linewidth]{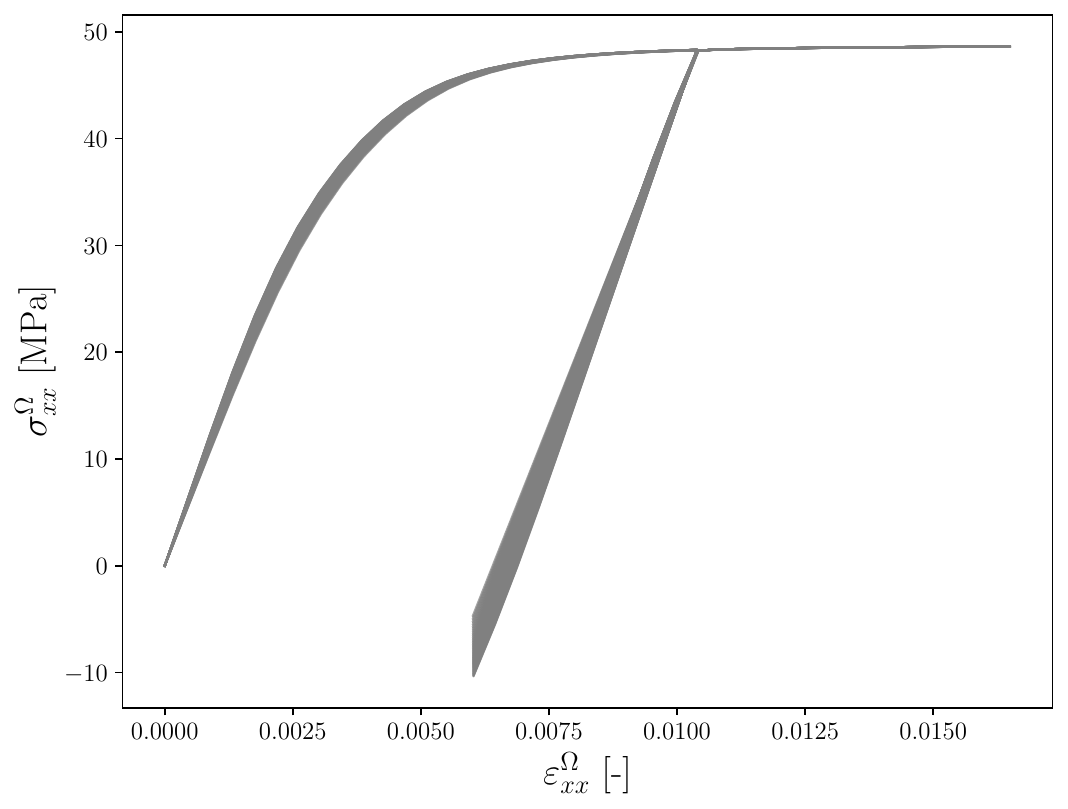} 
        \caption{Young's modulus of fiber}
        \label{ef}
    \end{subfigure}
    \hfill
    \begin{subfigure}{0.49\textwidth}
        \centering
        \includegraphics[width=\linewidth]{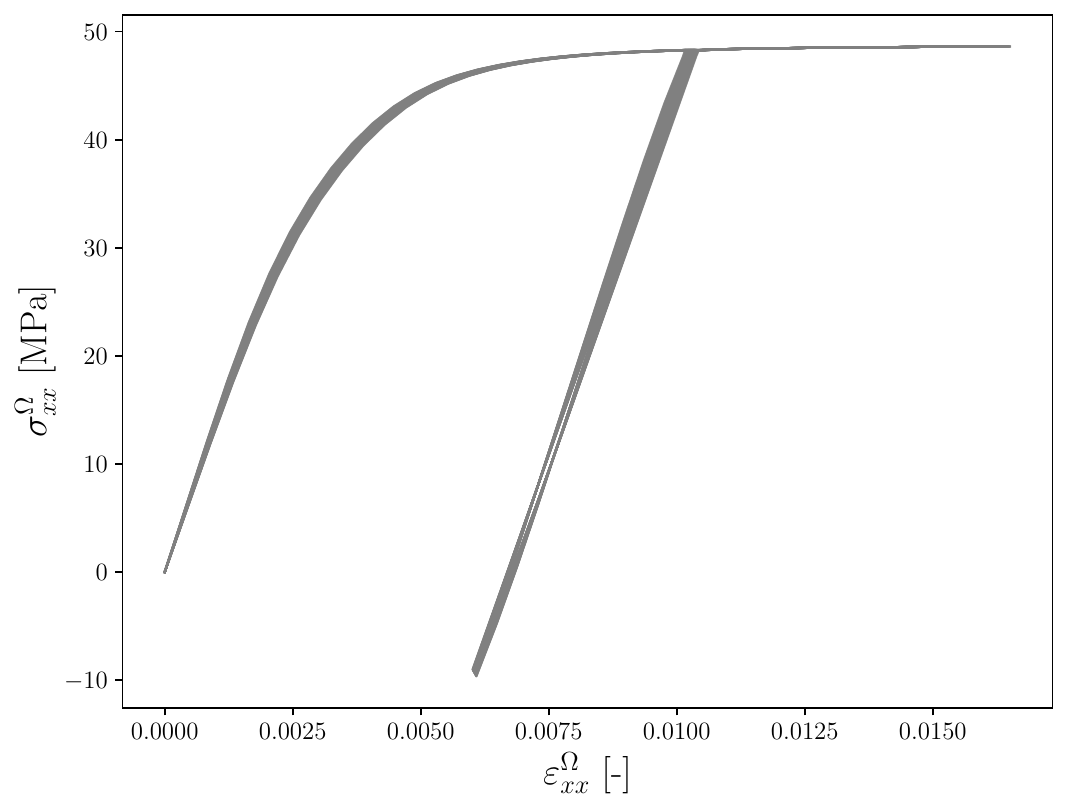}
        \caption{Poisson's ratio of fiber}
        \label{nf}
    \end{subfigure}
    \end{minipage}
    \caption{Influence of varying material properties on microstructural response}
    \label{fig:main}
\end{figure}

To evaluate how well the PRNN generalizes for microscale responses with properties different from the ones used for training it, micromodels are generated by varying one parameter at a time across the following intervals: $E_m \in [200, 8000]$ MPa, $E_f \in [40000, 100000]$ MPa, $\nu_m \in [0.1, 0.7]$, $\nu_f \in [0.1, 1.0]$, and $\sigma_y \in [49, 109]$ MPa.\footnote{Note that in 2D plane stress analysis, the Poisson’s ratio can take values up to 1.0 without leading to singular behavior.} For the yield strength, only the first parameter of the yield function defined in Equation \ref{eq:y} is modified. Note that these intervals are chosen to explore the limitations of the model, if any, and are wider than typical ranges encountered in polymer composites. Each resulting micromodel is loaded in the same 10 predefined random directions to reduce variability due to different load paths and highlight the trends created by the various material properties. The stress-strain curves are shown in Figure \ref{fig:main} for one of the predefined load paths to illustrate the microstructural sensitivity to parameter variations. Figures \ref{em} and \ref{plc} highlight the significant influence of the Young's modulus ($E_m$) and yield strength ($\sigma_y$) of the matrix on the microstructural response, indicating strong parameter sensitivity. In contrast, Figure \ref{nm} shows less variation due to the change of the Poisson's ratio of the matrix ($\nu_m$). Similarly, Figures \ref{ef} and \ref{nf} suggest that variations in fiber properties have minimal impact on the response of the micromodel for this geometry. Therefore, in this work, the effect of varying $E_m$ and $\sigma_y$ is studied. The low sensitivity of the micromodel to variations in the other parameters indicates that the pretrained PRNN, with fixed values for these parameters will remain accurate in the intervals considered.

Two approaches can be employed when testing the pretrained PRNN on curves obtained from the microstructures with varying material properties. The parameters in the material layer can either be fixed to the values used during training, or can be adjusted to match the properties of the input curve. In the latter approach, the properties in the material model embedded in the material points of the PRNN are dynamically adjusted for each input curve resulting from that set of parameters. From now on, the following notations are used to differentiate between these two approaches: PRNN$_f$ for fixed properties, and PRNN$_v$ for varying properties during testing. It is important to note that the weights of the trained PRNN are optimized for the fixed set of material properties of Section 2.4 and are not modified.

To compare the results, a relative error is used to account for the different stress levels resulting from the variation of material properties. The relative error is computed by dividing the absolute mean squared error by the absolute value of the maximum stress achieved in the micromodel.

\subsection{Varying the Young's modulus of the matrix}

\begin{figure}[!h]
\centering
\includegraphics[scale=0.45]{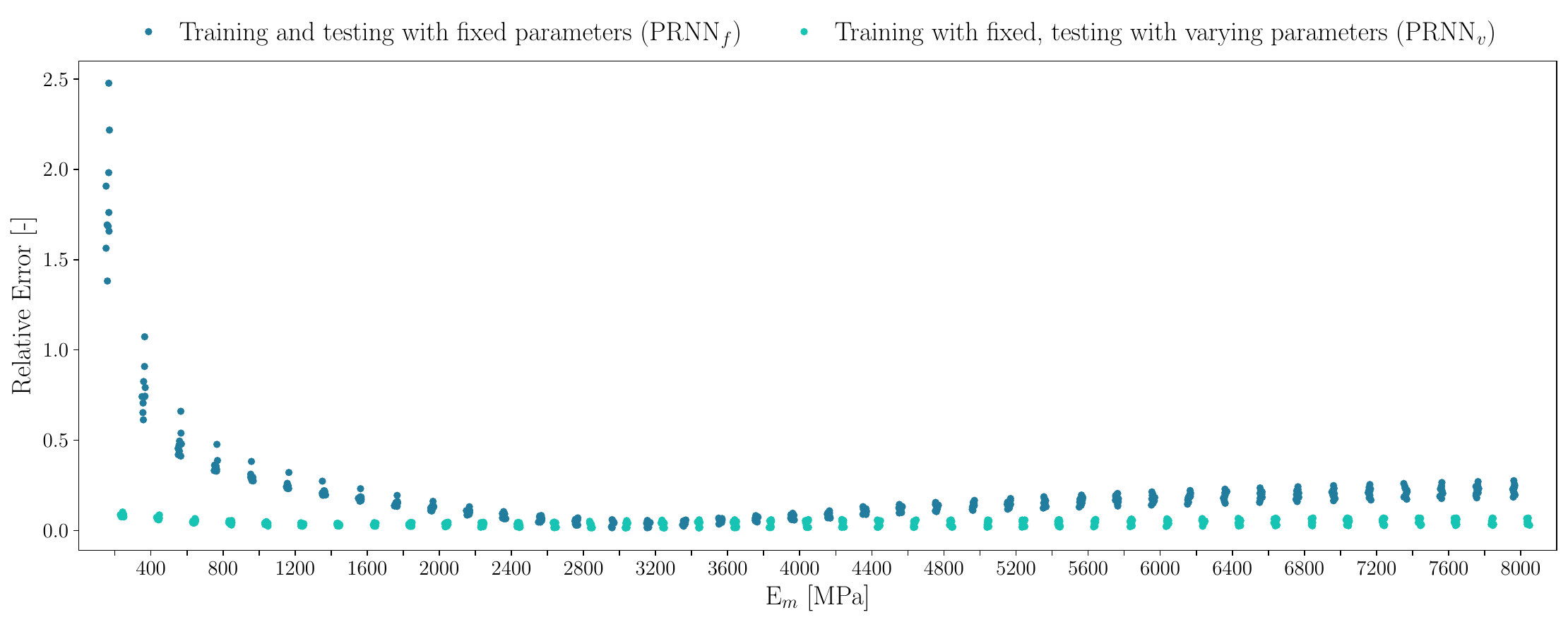}
\caption{\centering Test error for PRNN$_f$ and PRNN$_v$ across the selected E$_m$ values}
\label{fig:emall}
\end{figure}

\begin{figure}[!h]
  \centering
  \begin{subfigure}[b]{0.45\textwidth}
    \centering
    \includegraphics[width=\textwidth]{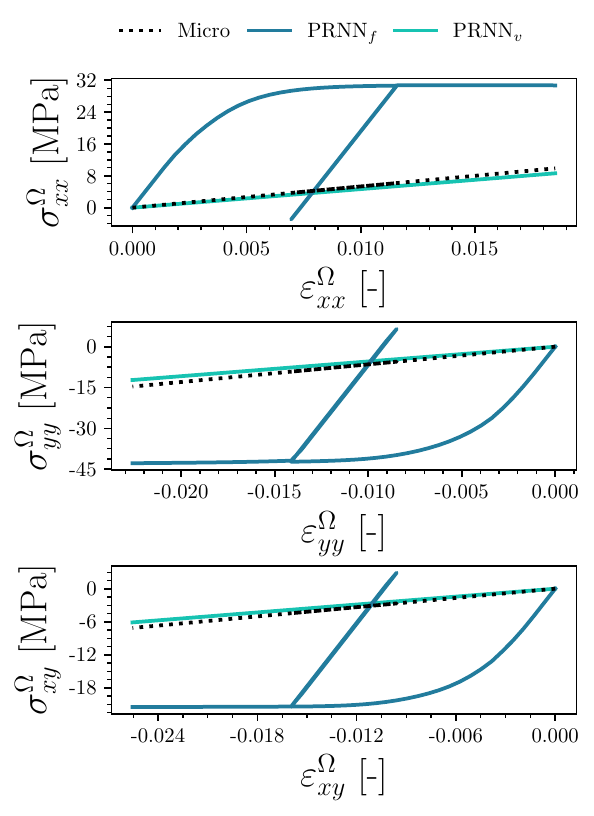}
    \caption{\centering Prediction on a microstructure response with E$_m = 200$ MPa}
    \label{fig:1ms}
  \end{subfigure}
  \hfill
  \begin{subfigure}[b]{0.45\textwidth}
    \centering
    \includegraphics[width=\textwidth]{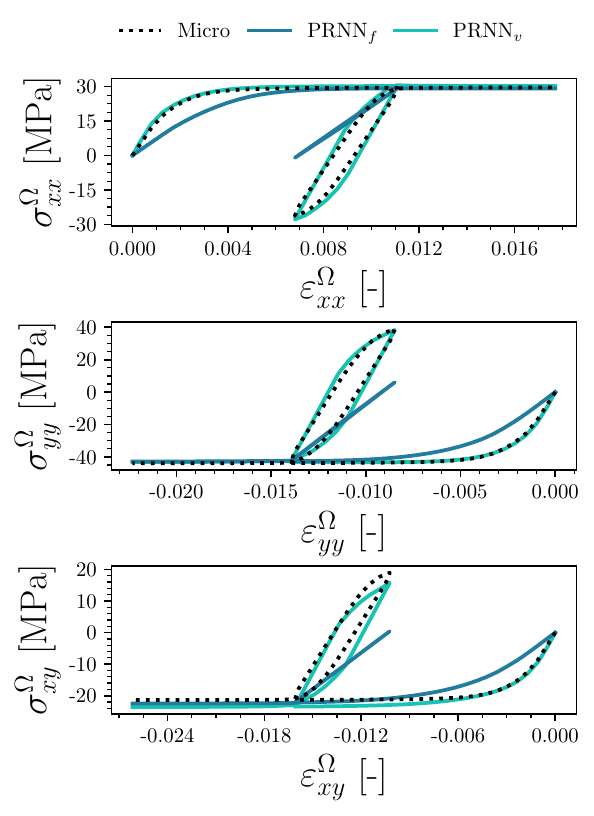}
    \caption{\centering Prediction on a microstructure response with E$_m = 8000$ MPa}
    \label{fig:1lc}
  \end{subfigure}
\caption{\centering Prediction of PRNN$_f$ and PRNN$_v$ on test curves with different E$_m$ values}
  \label{fig:p1}
\end{figure}

First, the performance of PRNN$_f$ and PRNN$_v$ is evaluated on microstructure responses with different $E_m$ values. The relative MSE is shown in Figure \ref{fig:emall} for each test curve across the selected $E_m$ values for both approaches. The quality of the predictions from the two surrogate strategies is illustrated in Figure \ref{fig:p1}, where the prediction of PRNN$_f$ and PRNN$_v$ is compared with the response of the micromodel for two values of $E_m$.

PRNN$_f$ performs well within a narrow range around the training parameter ($E_m = 3130 MPa$) but loses accuracy as $E_m$ deviates from this range, with highest relative errors at low $E_m$ values. The micromodel has a linear elastic response for these low values, which the network with a higher modulus in its material model cannot capture. Instead, for input strains where the micromodel remains elastic, the network predicts a response that enters the plastic regime. This behavior is illustrated in Figure \ref{fig:1ms}. At higher $E_m$ values, although Figure \ref{fig:emall} shows a smaller increase in relative error, Figure \ref{fig:1lc} reveals that the homogenized stiffness prediction is as inaccurate as for low $E_m$ values.



On the other hand, PRNN$_v$ predicts with consistently low relative error across the whole range of $E_m$ values, highlighting its ability to extrapolate well to unseen $E_m$ values. In addition to its low errors, Figure \ref{fig:p1} shows that unlike PRNN$_f$, the homogenized linear elastic stiffness prediction is also captured accurately.

\subsection{Varying the yield strength of the matrix}

\begin{figure}[!h]
\centering
\includegraphics[scale=0.45]{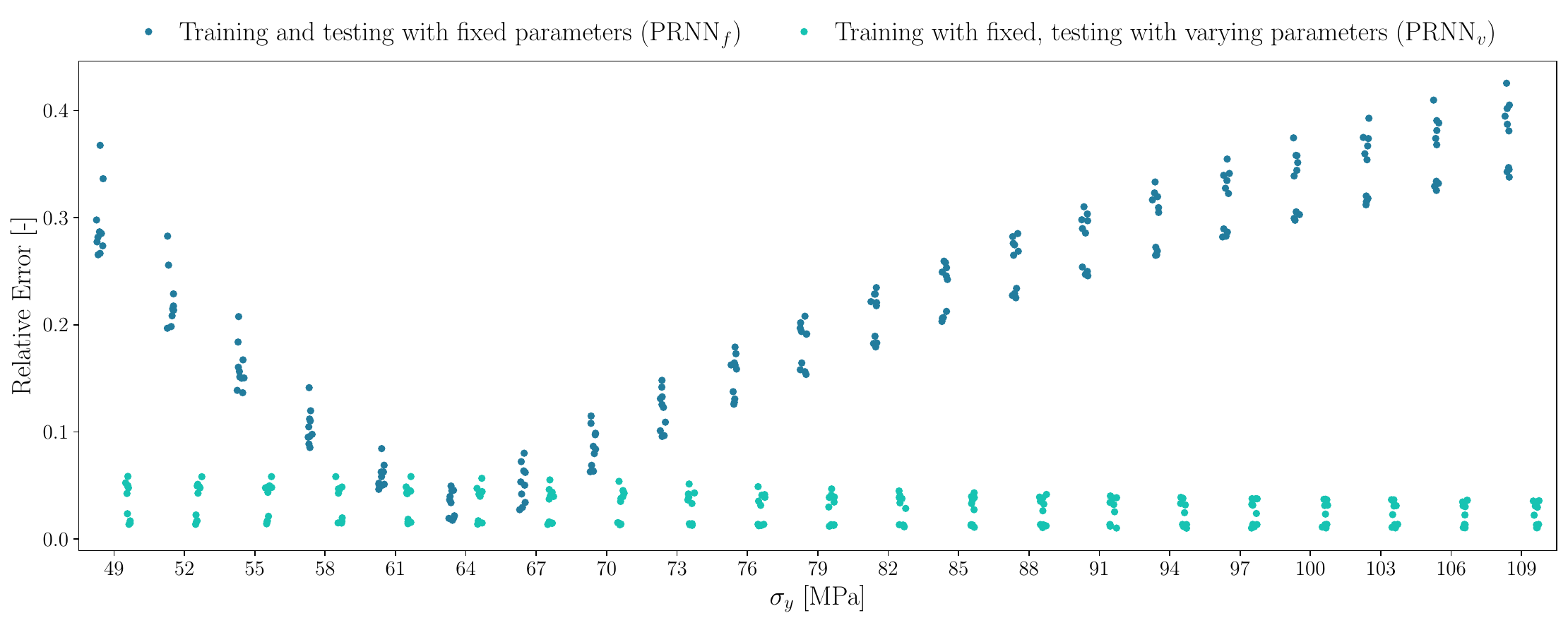}
\caption{\centering Test error for PRNN$_f$ and PRNN$_v$ across the selected $\sigma_y$ values}
\label{fig:plcall}
\end{figure}

\begin{figure}[!ht]
\minipage{0.49\textwidth}
  \centering
  \includegraphics[width=0.9\linewidth]{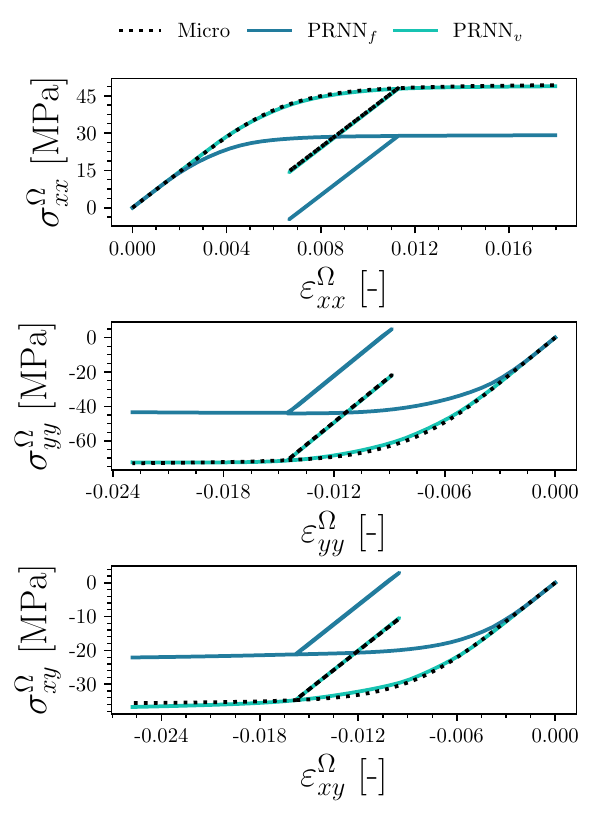}
  \caption{\centering Prediction of PRNN$_f$ and PRNN$_v$ on a microstructure response with $\sigma_y = 109$ MPa}\label{fig:plch}
\endminipage\hfill
\minipage{0.49\textwidth}
  \centering
  \includegraphics[width=0.9\linewidth]{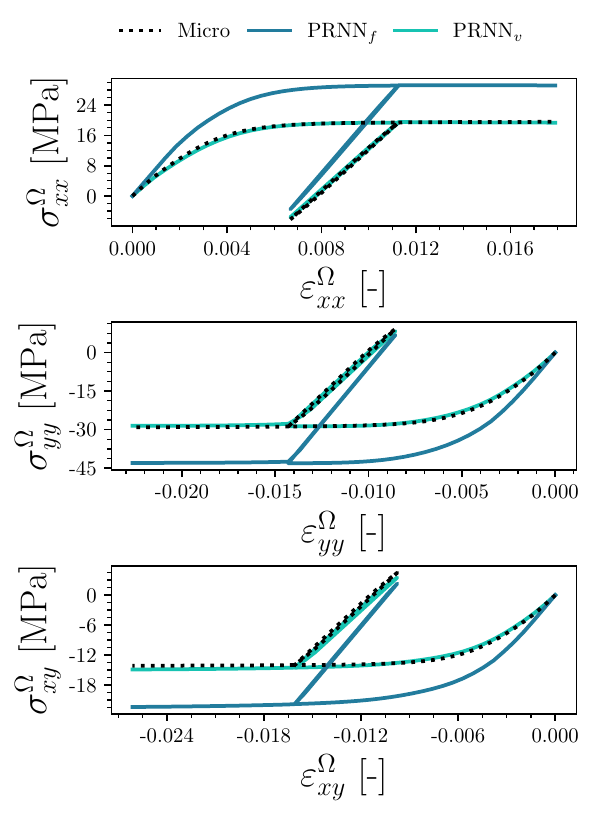}
  \caption{\centering Prediction of PRNN$_f$ and PRNN$_v$ on a microstructure response with E$_m = 2200$ MPa and $\sigma_y = 40$ MPa }\label{fig:p5}
\endminipage\hfill
\end{figure}

The predictions of PRNN$_f$ and PRNN$_v$ are next evaluated on curves resulting from microstructures with varying $\sigma_y$ values. The relative MSE on each test curve is shown in Figure \ref{fig:plcall} for both approaches. Similar to the analysis on $E_m$, PRNN$_f$ predicts accurately only within a small interval around the training parameter of $\sigma_y = 64.8$ MPa, while PRNN$_v$ is able to capture the response with low error across the whole range. There is a more pronounced increase in relative MSE for PRNN$_f$ as $\sigma_y$ deviates further from the training value. The network either over- or underpredicts the stress values, depending on the parameter's value. For illustration, Figure \ref{fig:plch} shows the case of $\sigma_y = 109 MPa$ where PRNN$_f$ underpredicts the stress response, while PRNN$_v$ matches the full-order response with high accuracy.

\subsection{Varying both the Young's modulus and yield strength of the matrix}

Previously, when the studied parameters are varied one at a time, PRNN$_v$ significantly outperformed PRNN$_f$, showing an excellent accuracy. Therefore, next we investigate whether this trend holds when both parameters in the micromodel are varied at the same time. For this analysis, the PRNNs are evaluated on a grid of parameter combinations. It is important to note that in some cases, specifically for micromodels with high Young's modulus and low yield strength, generating the full-order solution is not possible due to model convergence issues.

\begin{figure}[htbp]
    \centering
    \begin{minipage}{\textwidth}
    \centering
    \begin{subfigure}{0.48\textwidth}
        \centering
        \includegraphics[width=\linewidth]{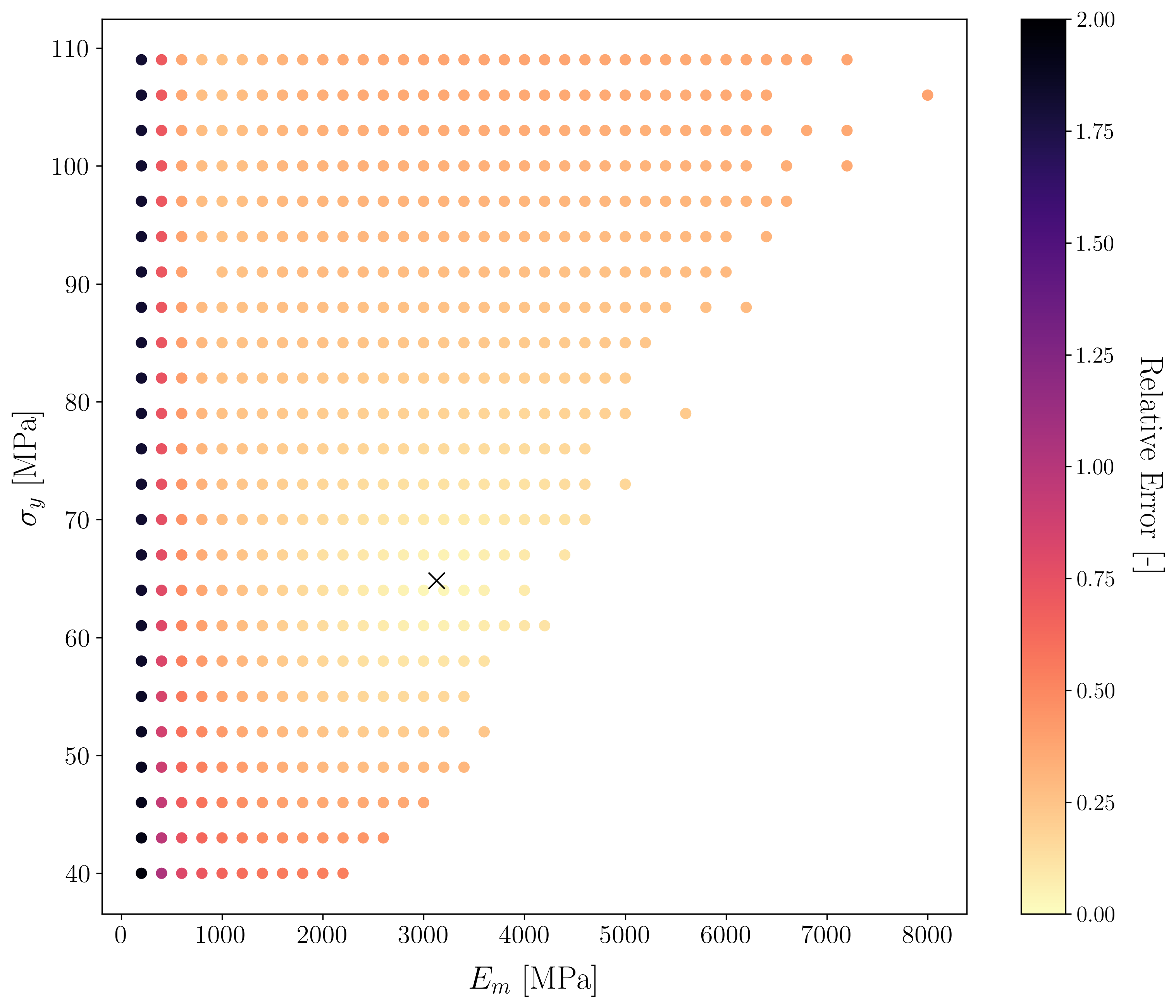}
        \caption{Test error for PRNN$_f$}
        \label{fig:3ms}
    \end{subfigure}
    \hfill
    \begin{subfigure}{0.48\textwidth}
        \centering
        \includegraphics[width=\linewidth]{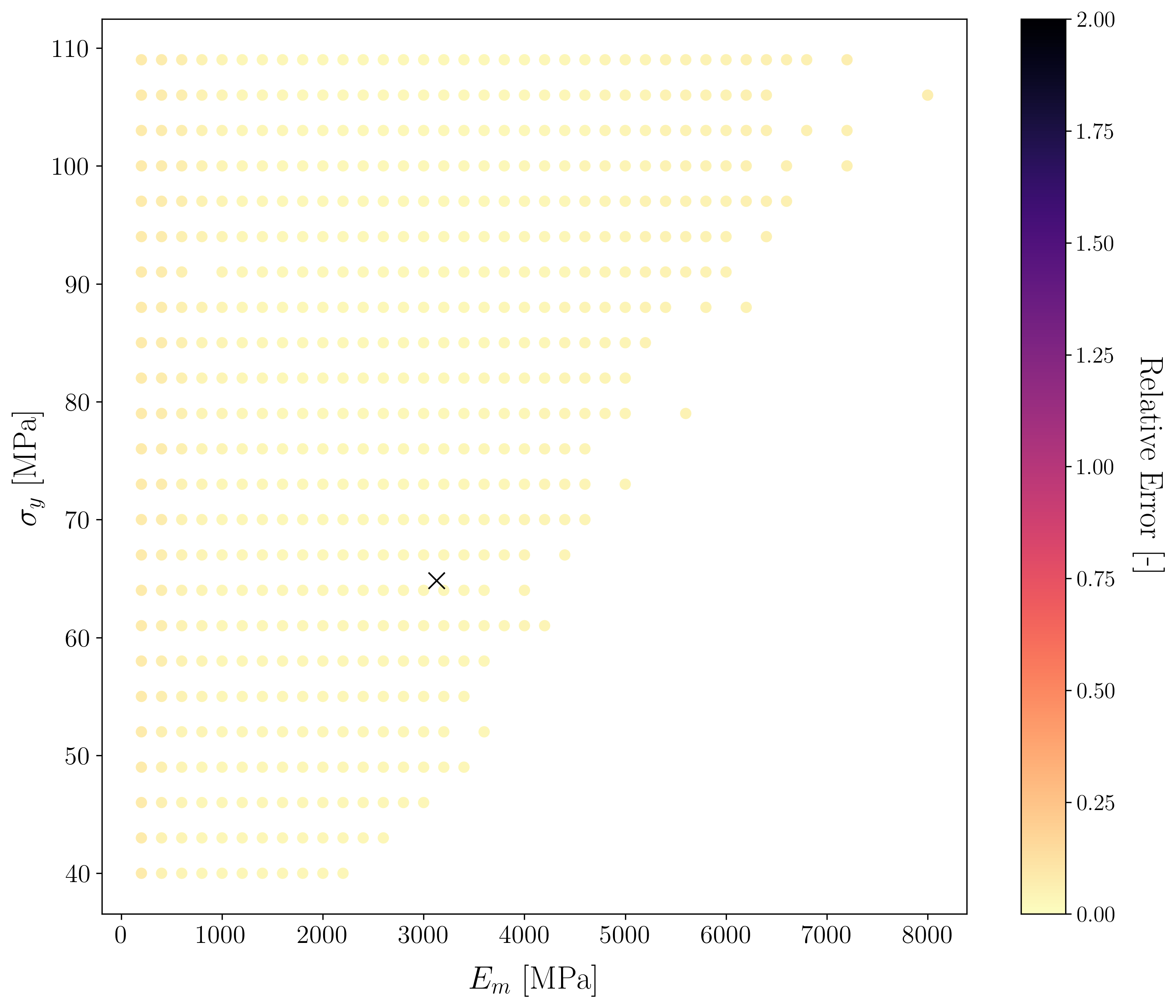}
        \caption{Test error for PRNN$_v$}
        \label{fig:3lc}
    \end{subfigure}
    \end{minipage}
    \caption{Test error for PRNN$_f$ and PRNN$_v$ for combinations of the selected E$_m$ and $\sigma_y$ values}
    \label{fig:34}
\end{figure}

The mean values of the relative MSE on the test curves are shown in Figures \ref{fig:3ms} and \ref{fig:3lc} for PRNN$_f$ and PRNN$_v$, respectively. The training parameters are marked with a black cross for reference. Figure \ref{fig:3ms} illustrates that PRNN$_f$ predicts accurately only within a small radius around the training parameters, which agrees well with the trends observed when only one parameter was varied.

Although PRNN$_v$ performs significantly better than PRNN$_f$, a slight difference in material stiffness between the full-order model and the prediction of PRNN$_v$ appears. Since the micromodel behavior is primarily elastic for low $E_m$ values, this effect of the small difference in stiffness becomes more visible in this region, which is represented in Figure \ref{fig:3lc} with light orange colors at the left edge. However, the overall performance of PRNN$_v$ remains accurate for the given parameter intervals.

Figure \ref{fig:p5} shows predictions on a representative curve from the micromodel with $E_m = 2200$ MPa and $\sigma_y = 40$ MPa. PRNN$_f$ overshoots the stress level due to the decreased yield strength and overpredicts the stiffness resulting from the change in Young's modulus of the matrix. PRNN$_v$ on the other hand is able to adapt to the response of the micromodel, further showcasing its predictive capability across different combinations of material properties.


The performance of PRNN$_f$ confirms the expected rigidity of regular neural networks. When trained on a fixed parameter set without incorporating physics-based information, conventional networks struggle with extrapolating to other material parameters. This aligns well with the assumption that a standard neural network would either require retraining each time a new parameter set is considered, or additional material parameter-related inputs during training. As a result, accounting for uncertainties in the micromodel coming from material properties becomes an arduous task. On the other hand, PRNN$_v$ predicts with high accuracy both when properties are varied one at a time, or as a combination. This enables the network to account for uncertainties originating from material properties without requiring otherwise extensive retraining.



\section{PRNN-driven multiscale uncertainty quantification}
\label{}
Given the accurate predictions of PRNN$_v$ on the microscale for varying parameters, its performance in a multiscale setting is evaluated next. Previously, the PRNN was shown to be efficient in a multiscale framework when material properties are fixed \cite{MAIA2023115934}. Now we extend it to scenarios where the microscale material properties vary as described in the previous section to evaluate whether its performance on the macroscale is maintained.

\subsection{Uncertainty quantification on a coarse mesh}

\begin{figure}[!h]
\centering
\includegraphics[scale=0.35]{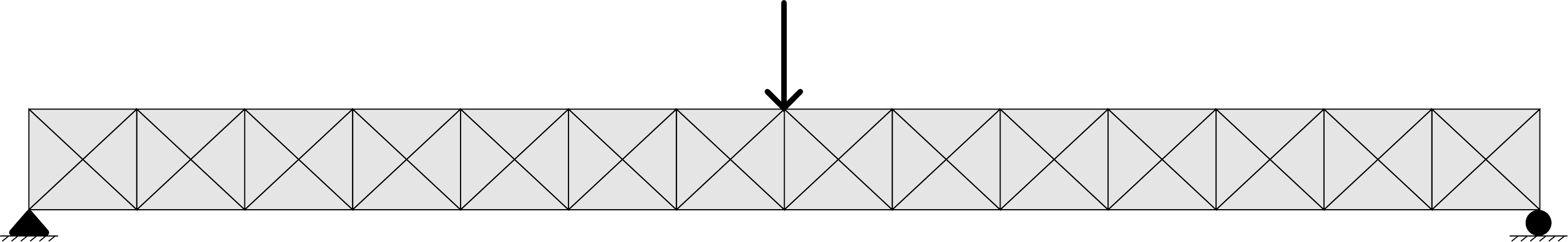}
\caption{\centering Macroscale model of simply supported beam with coarse mesh}
\label{fig:tpb}
\end{figure}

For this study, multiscale uncertainty quantification is performed on a simple three-point bending test with varying material properties. It is assumed that microscopic material parameters are uniform over the domain but that their value is uncertain according to known distributions. A displacement of 4 mm is applied in 80 steps to the midpoint of a simply supported beam with length of 120 mm and height of 8 mm, and the forces at midspan are computed over time. For microscale material variability, 5000 parameters are sampled with uniform distribution from the intervals of Young's modulus and yield strength of the matrix, specified in Section 3. The load distributions resulting from the full-order model for different levels of applied displacement are compared to those obtained with the PRNN$_v$-driven model. Due to the high computational cost associated with full-order simulations, an overly coarse mesh with 56 elements (shown in Figure \ref{fig:tpb}) is considered for this comparison, with each full order simulation taking approximately 2 hours. This allows for a fair comparison between the full-order model and the PRNN$_v$-driven model because both will have the same bias due to macroscale discretization error, while the microscale histories will remain representative for the actual structure.

\begin{figure}[!h]
  \centering
  \begin{subfigure}[b]{0.325\textwidth}
    \centering
    \includegraphics[width=\textwidth]{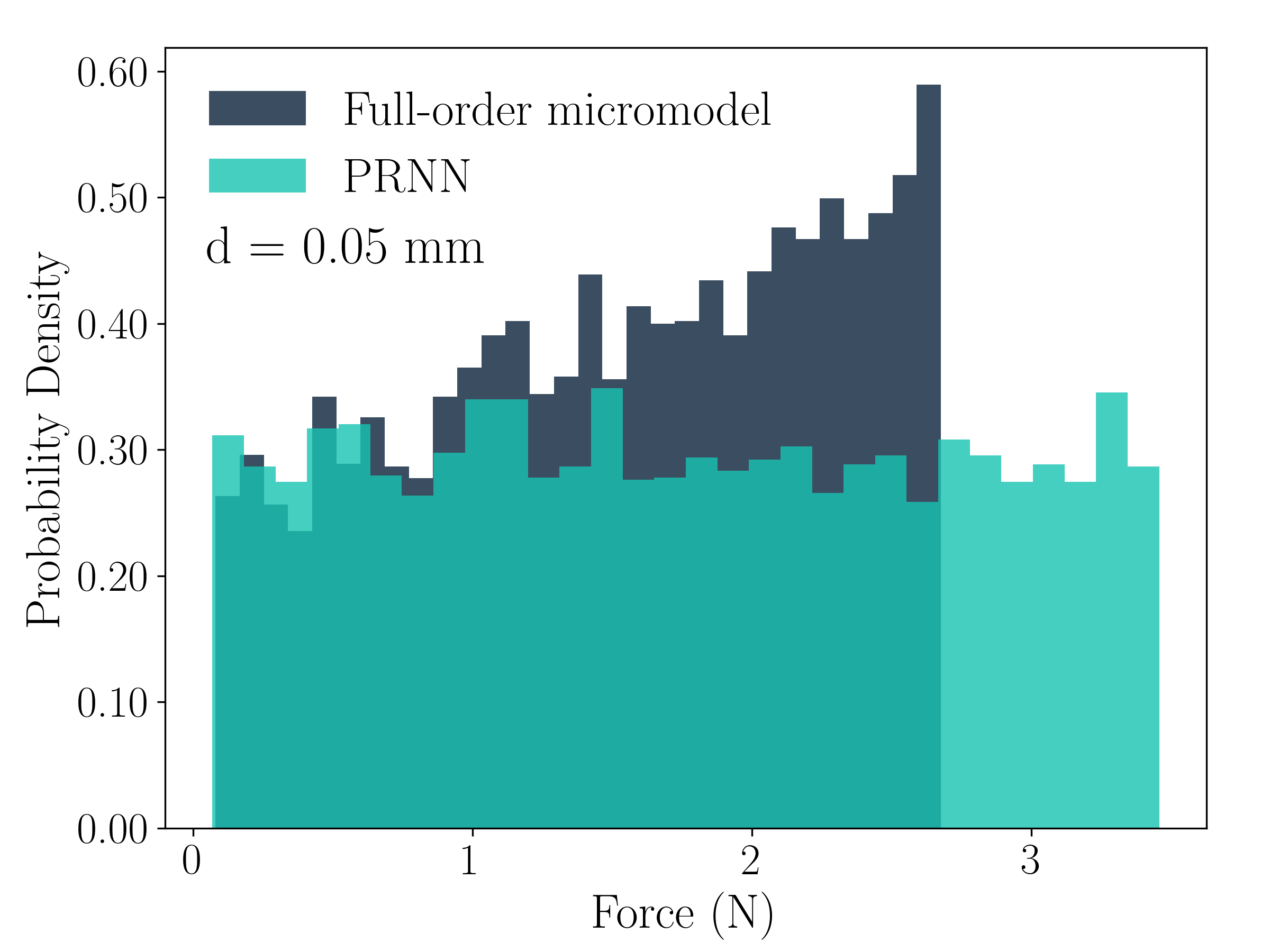}
    \caption{\centering Displacement = 0.05 mm}
    \label{fig:nnhfem1}
  \end{subfigure}
  \hfill
  \begin{subfigure}[b]{0.325\textwidth}
    \centering
    \includegraphics[width=\textwidth]{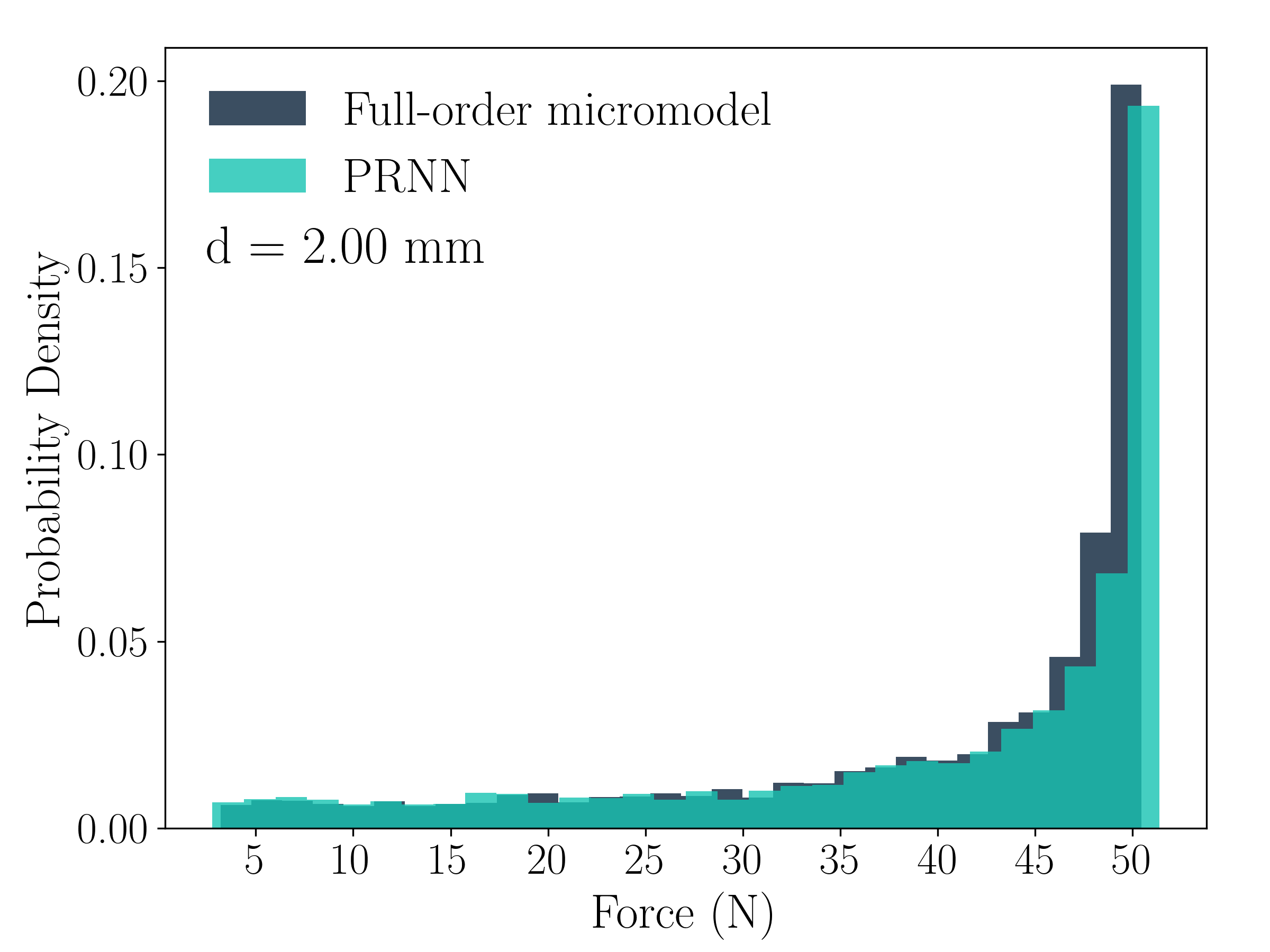}
    \caption{\centering Displacement = 2 mm}
    \label{fig:nnhfem2}
  \end{subfigure}
  \hfill
  \begin{subfigure}[b]{0.325\textwidth}
    \centering
    \includegraphics[width=\textwidth]{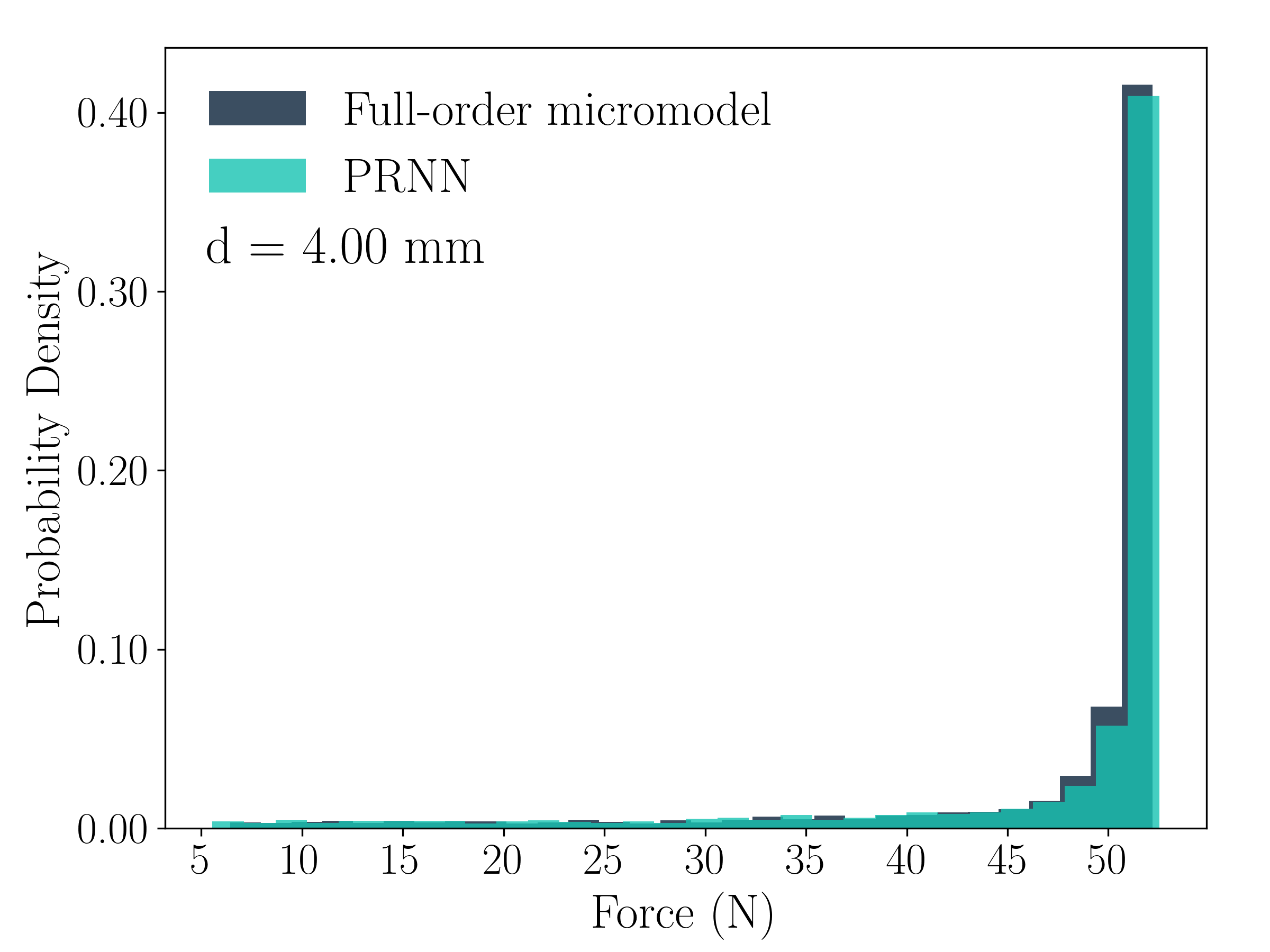}
    \caption{\centering Displacement = 4 mm}
    \label{fig:nnhfem3}
  \end{subfigure}
\caption{\centering Load distributions at the midpoint of the beam for varying E$_m$, using a coarse mesh}
  \label{fig:nnhfem}
\end{figure}

\begin{figure}[!h]
  \centering
  \begin{subfigure}[b]{0.32\textwidth}
    \centering
    \includegraphics[width=\textwidth]{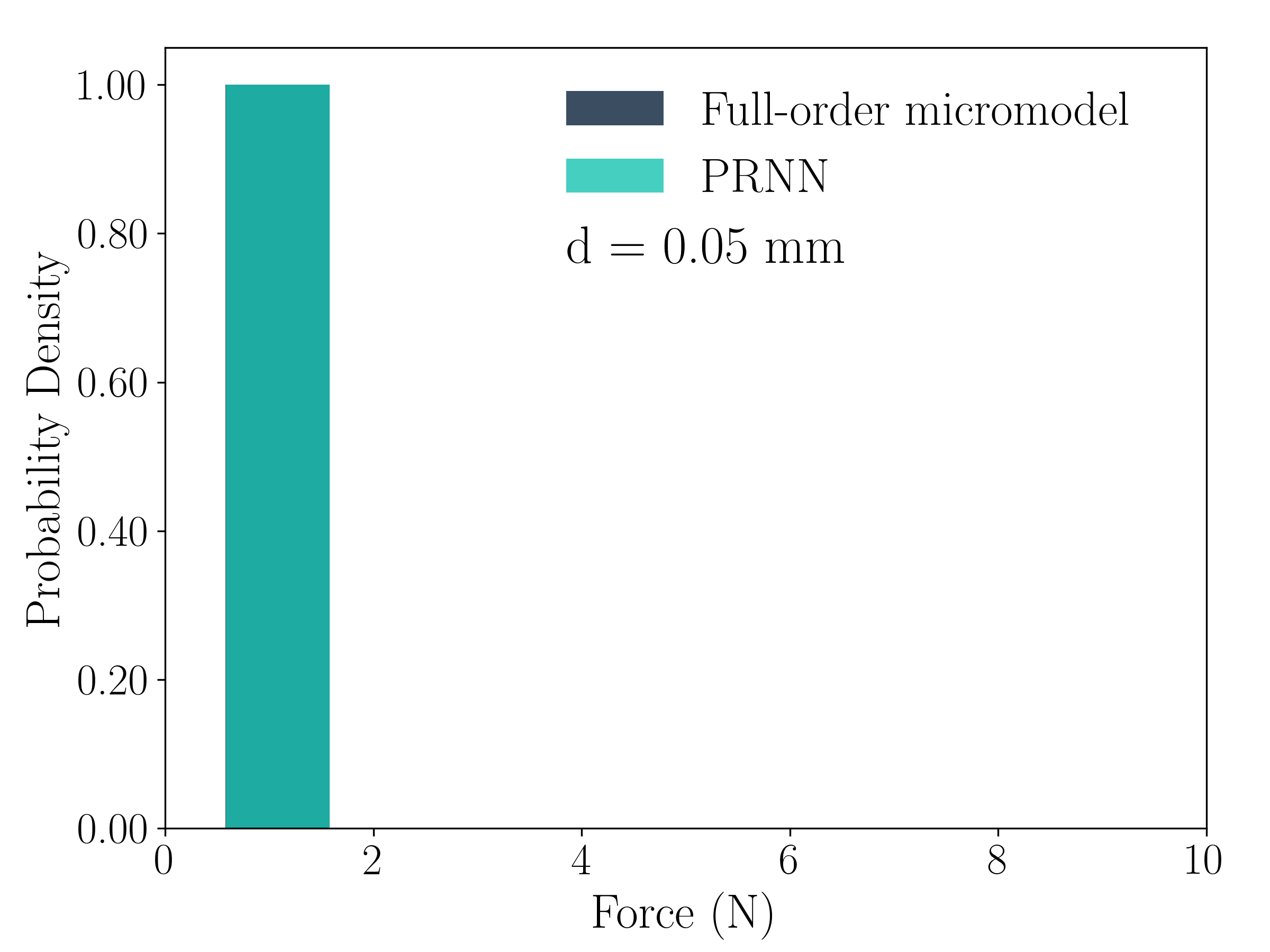}
    \caption{\centering Displacement = 0.05 mm}
    \label{fig:nnhfplc1}
  \end{subfigure}
  \hfill
  \begin{subfigure}[b]{0.32\textwidth}
    \centering
    \includegraphics[width=\textwidth]{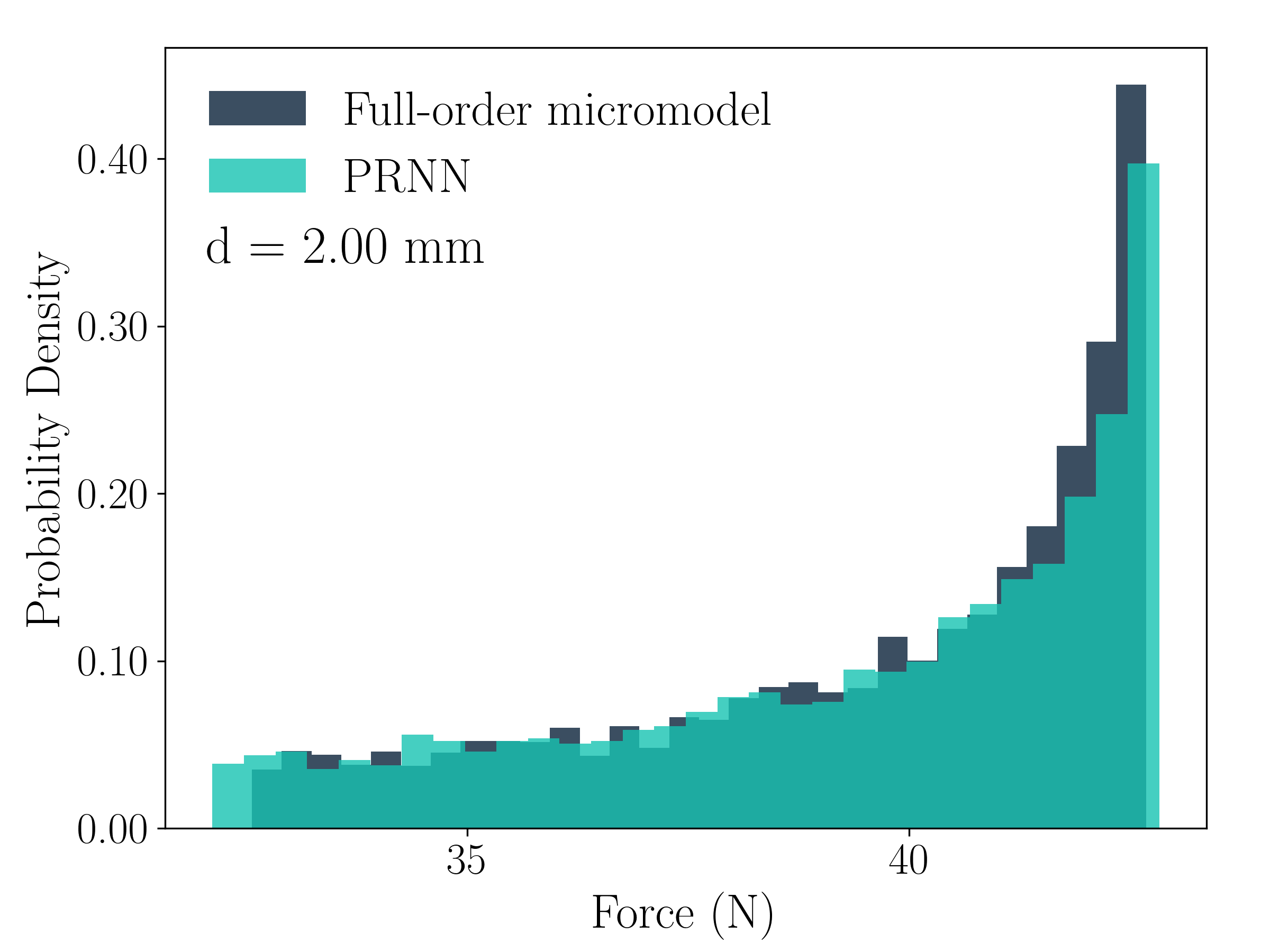}
    \caption{\centering Displacement = 2 mm}
    \label{fig:nnhfplc2}
  \end{subfigure}
  \hfill
  \begin{subfigure}[b]{0.32\textwidth}
    \centering
    \includegraphics[width=\textwidth]{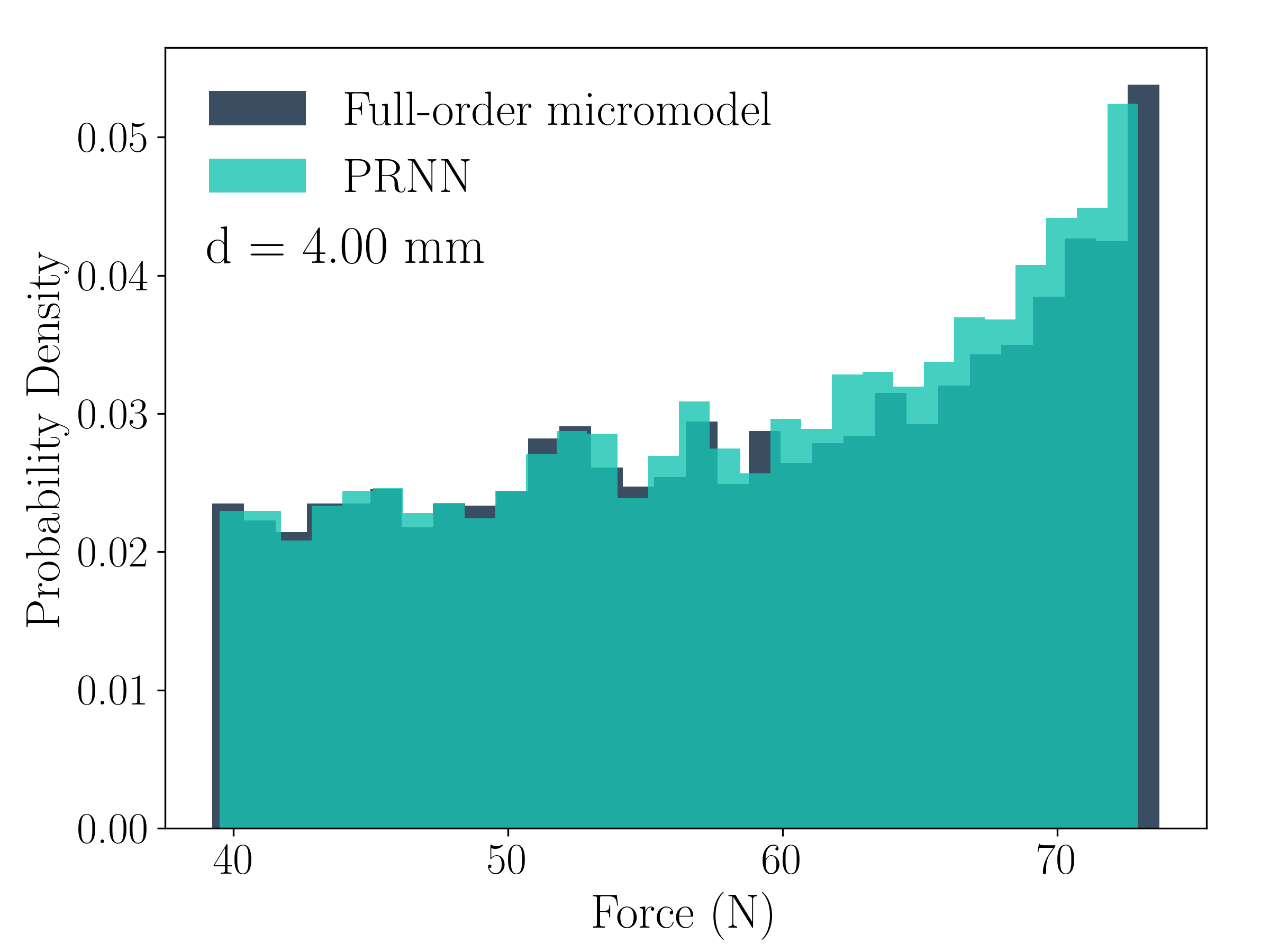}
    \caption{\centering Displacement = 4 mm}
    \label{fig:nnhfplc3}
  \end{subfigure}
\caption{\centering Load distributions at the midpoint of the beam for varying $\sigma_y$, using a coarse mesh}
  \label{fig:nnhfplc}
\end{figure}

\begin{figure}[!h]
  \centering
  \begin{subfigure}[b]{0.32\textwidth}
    \centering
    \includegraphics[width=\textwidth]{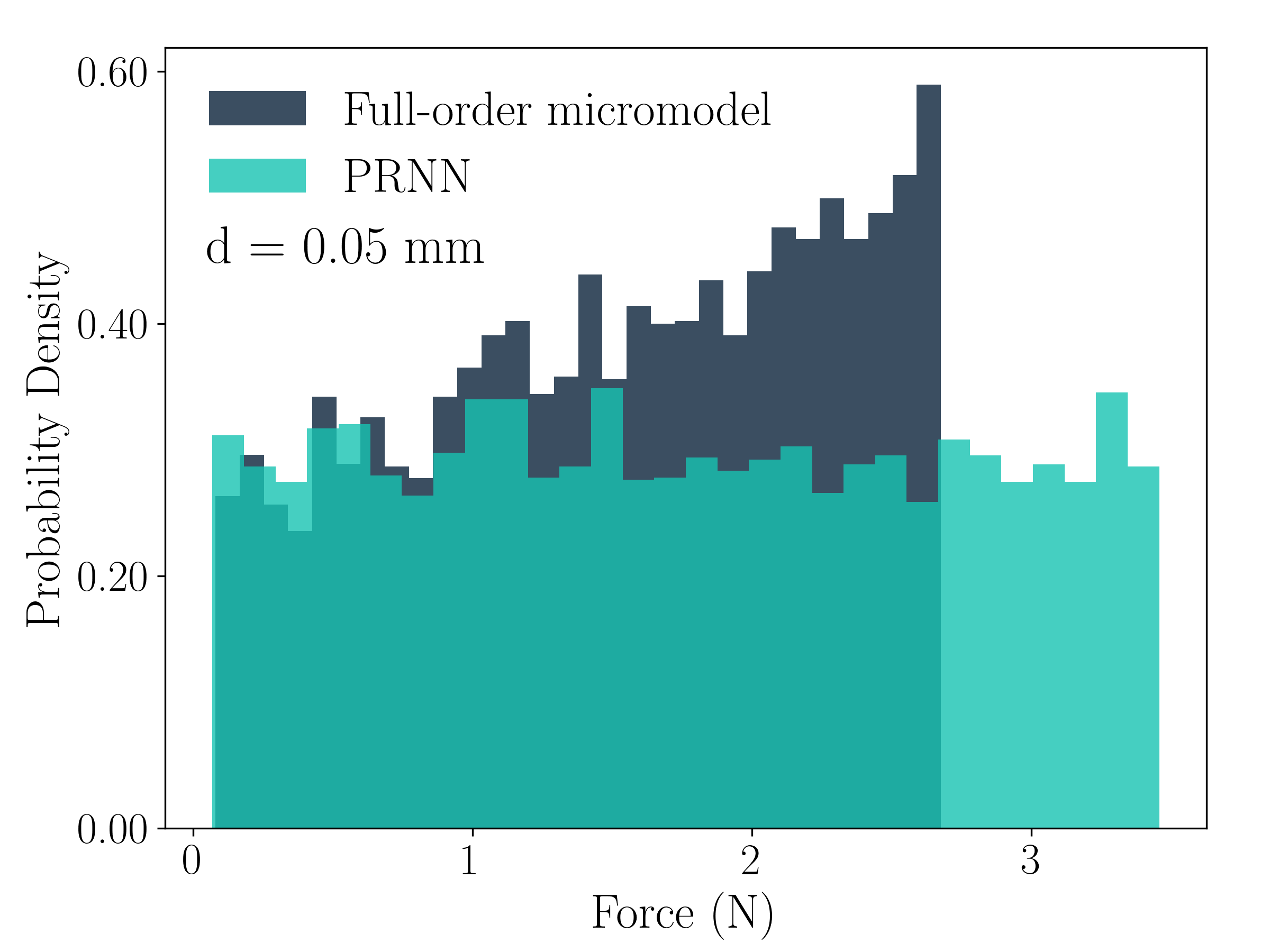}
    \caption{\centering Displacement = 0.05 mm}
    \label{fig:nnhfemplc1}
  \end{subfigure}
  \hfill
  \begin{subfigure}[b]{0.32\textwidth}
    \centering
    \includegraphics[width=\textwidth]{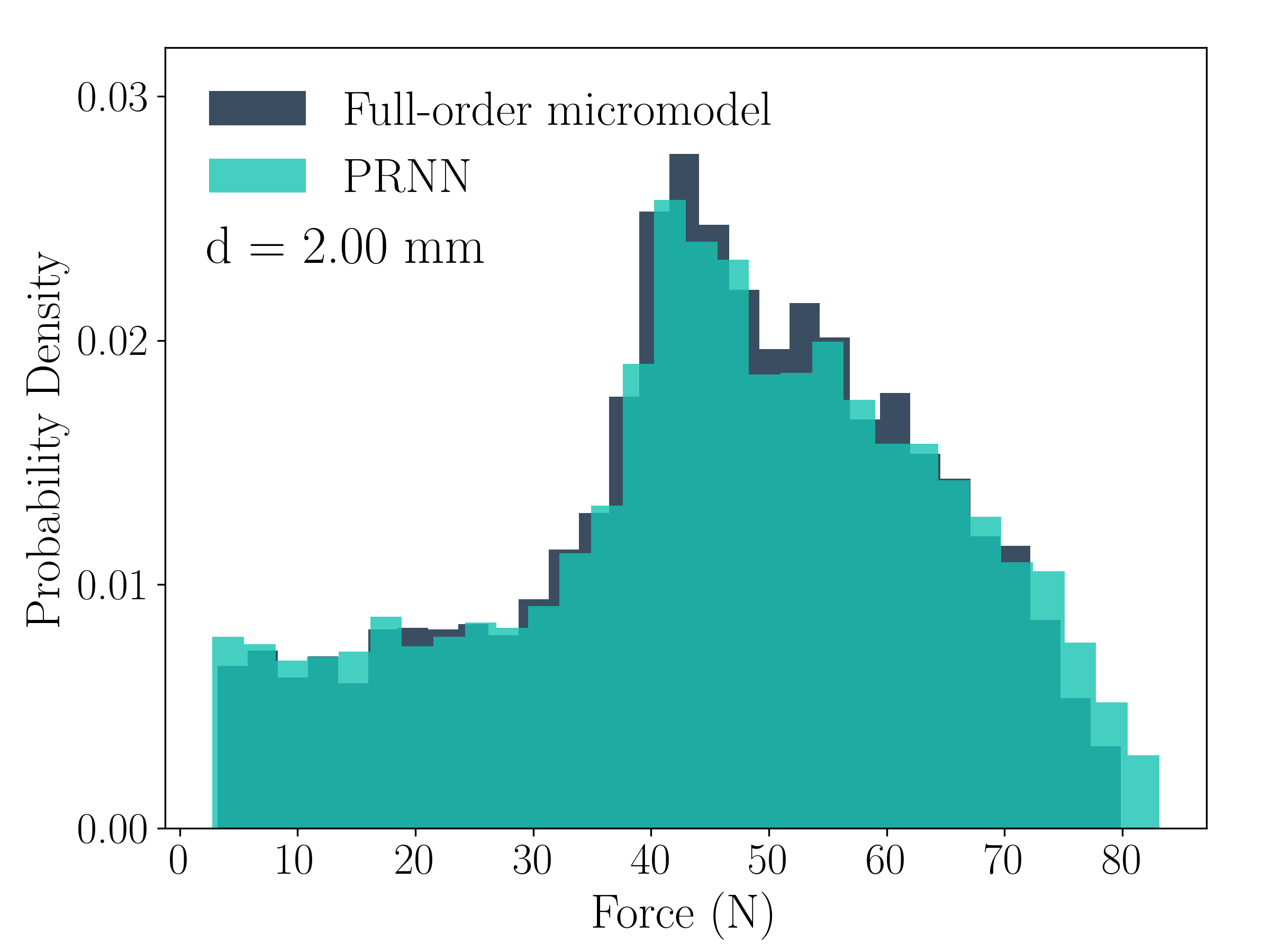}
    \caption{\centering Displacement = 2 mm}
    \label{fig:nnhfemplc2}
  \end{subfigure}
  \hfill
  \begin{subfigure}[b]{0.32\textwidth}
    \centering
    \includegraphics[width=\textwidth]{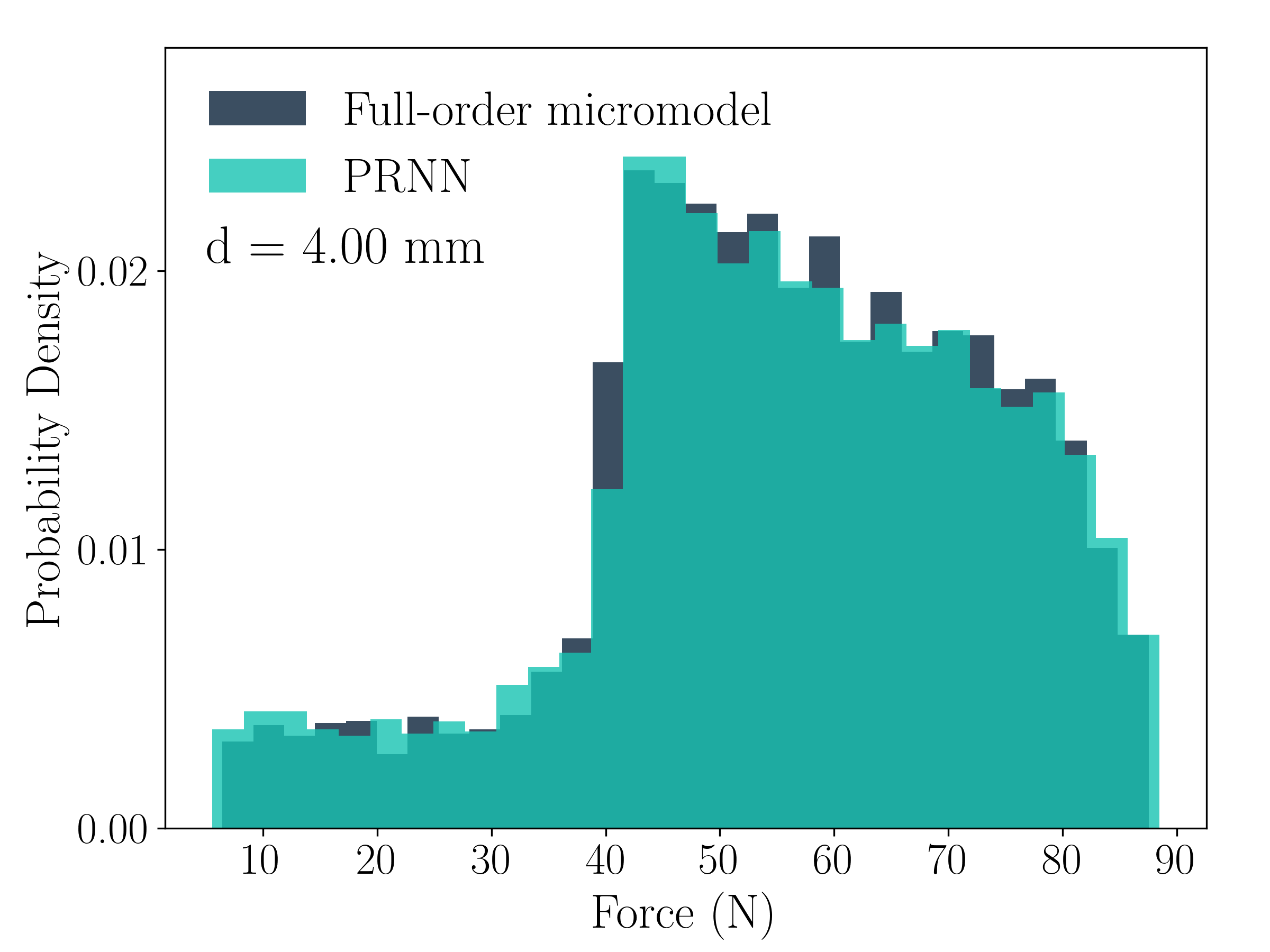}
    \caption{\centering Displacement = 4 mm}
    \label{fig:nnhfemplc3}
  \end{subfigure}
\caption{\centering Load distributions at the midpoint of the beam for varying E$_m$ and $\sigma_y$ simultaneously, using a coarse mesh}
  \label{fig:nnhfemplc}
\end{figure}

The force distribution histograms resulting from varying $E_m$ and $\sigma_y$, one at a time and simultaneously, are shown in Figures \ref{fig:nnhfem}-\ref{fig:nnhfemplc} at 3 different displacement levels. The histograms of the two methods overlap well, which shows that the PRNN-driven method accurately predicts the force distribution evolution over time.

\begin{figure}[!h]
  \centering
  \begin{subfigure}[b]{0.325\textwidth}
    \centering
    \includegraphics[width=\textwidth]{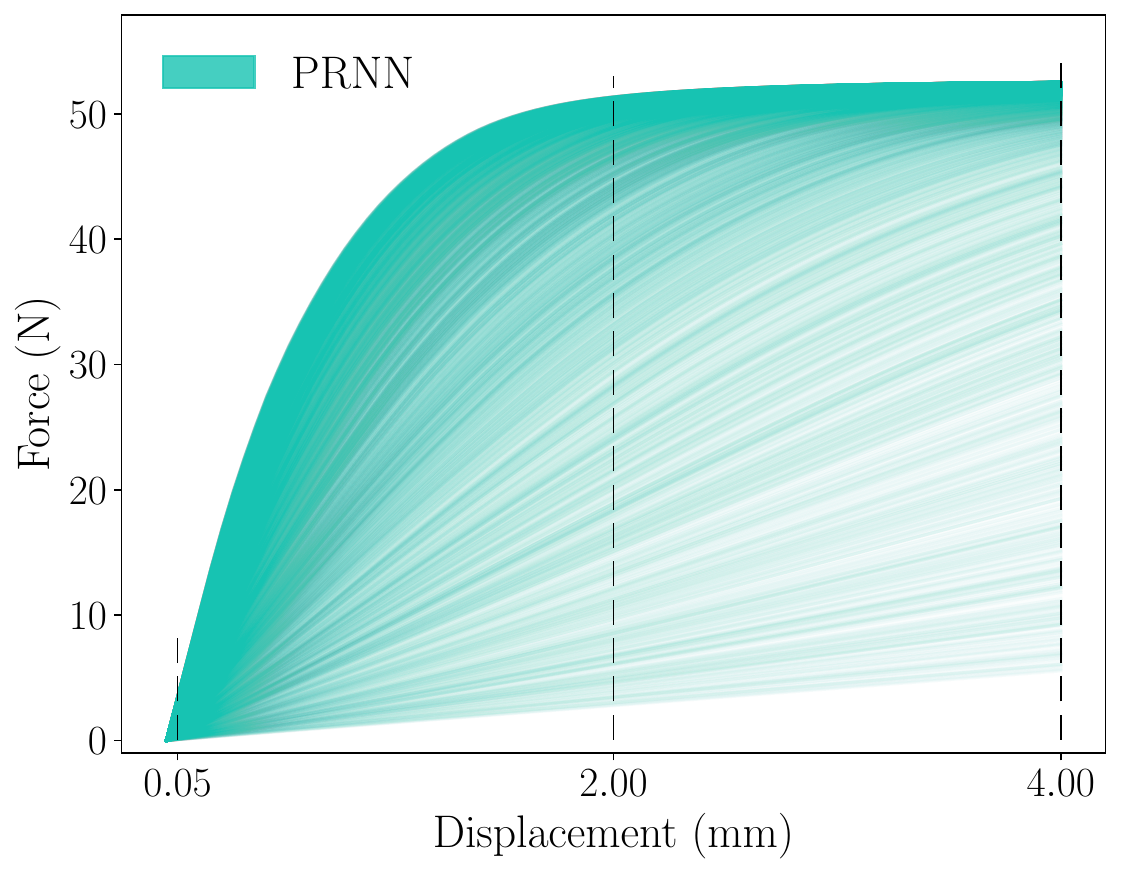}
    \caption{\centering Varying $E_m$}
    \label{fig:nnfdem}
  \end{subfigure}
  \hfill
  \begin{subfigure}[b]{0.325\textwidth}
    \centering
    \includegraphics[width=\textwidth]{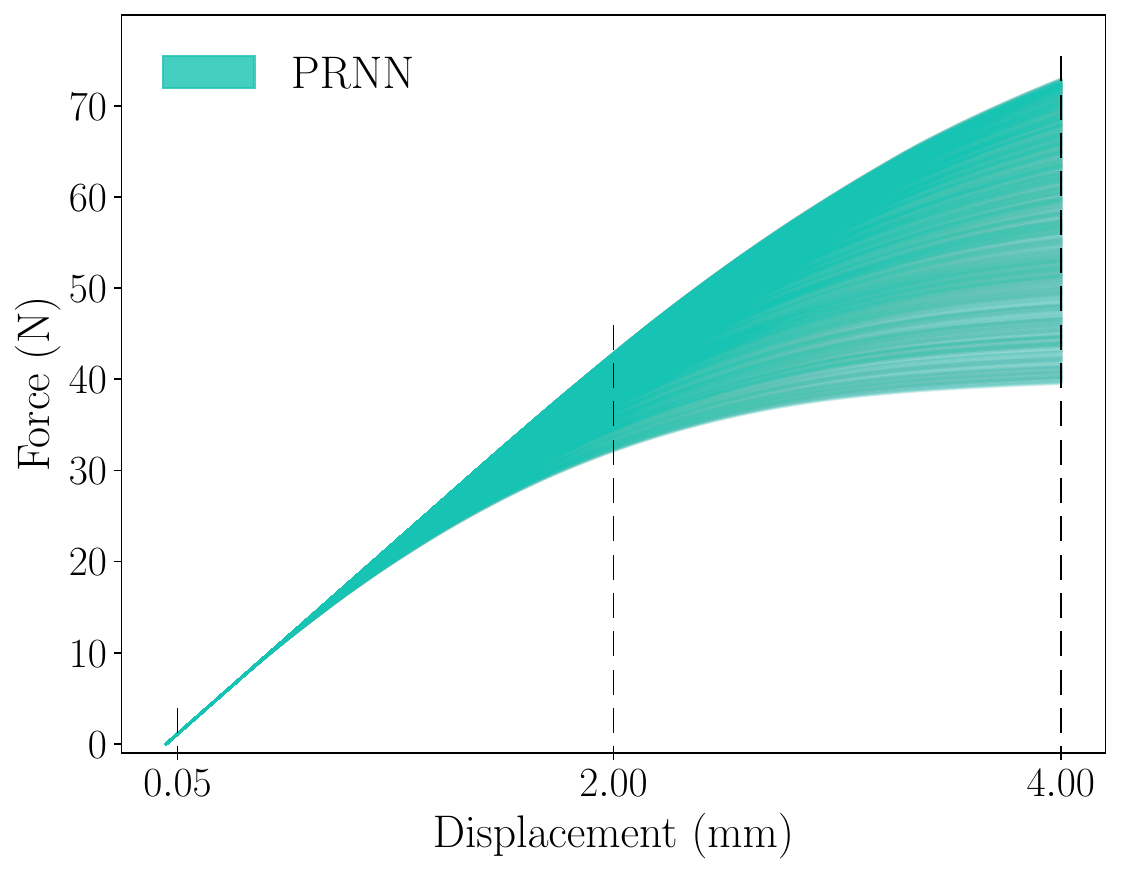}
    \caption{\centering Varying $\sigma_y$}
    \label{fig:nnfdplc}
  \end{subfigure}
  \hfill
  \begin{subfigure}[b]{0.325\textwidth}
    \centering
    \includegraphics[width=\textwidth]{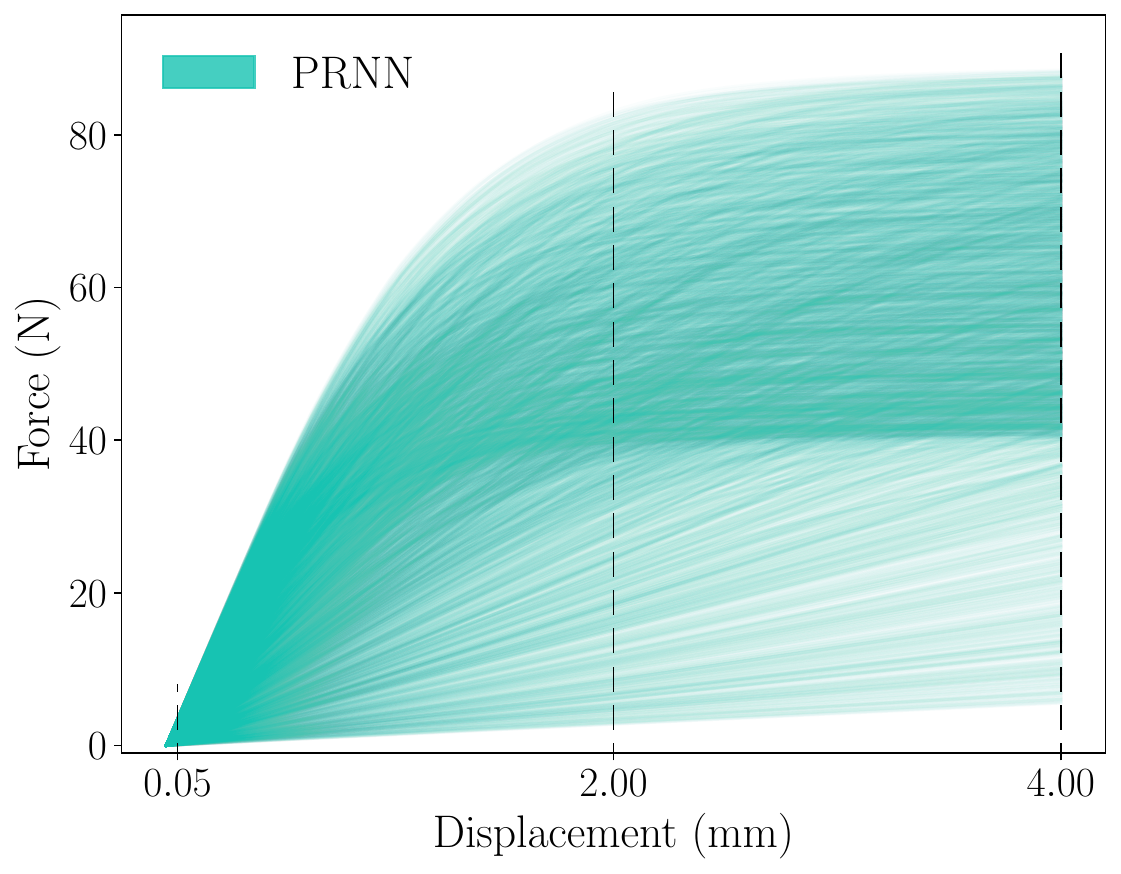}
    \caption{\centering Varying $E_m$ and $\sigma_y$}
    \label{fig:nnfdemplc}
  \end{subfigure}
\caption{\centering Force-displacement curves at the macroscale resulting from varying parameters at the microscale}
  \label{fig:nnfd}
\end{figure}

When only $E_m$ is varied, the PRNN$_v$-driven model is able to capture the transition from a more uniform force distribution to a highly skewed one (Figure \ref{fig:nnhfem}). The peak observed at $d = 2$mm and $d = 4$mm is explained by the corresponding force-displacement curves shown in Figure \ref{fig:nnfdem}. The increase in Young's modulus of the matrix causes more integration points to enter plasticity at earlier time steps, plateauing at the same load level. The PRNN-driven UQ also successfully predicts the force distributions when varying $\sigma_y$, transitioning from a single bar indicating the same force value across the parameter range, to a more uniform distribution. At small displacements, all the curves show the same linear elastic stiffness in Figure \ref{fig:nnfdplc} that explains the single bar. Finally, the highly nonlinear distribution resulting from the combined effect of varying the two parameters simultaneously is also well captured by using PRNN$_v$ (\ref{fig:nnhfemplc}). The higher density of curves around 40MPa and above at $d = 4$ mm, as seen in the curves in Figure \ref{fig:nnfdemplc}, explains the shape of the corresponding histogram in Figure \ref{fig:nnhfemplc3}.


\begin{figure}[htbp]
    \centering
    \includegraphics[scale=0.4]{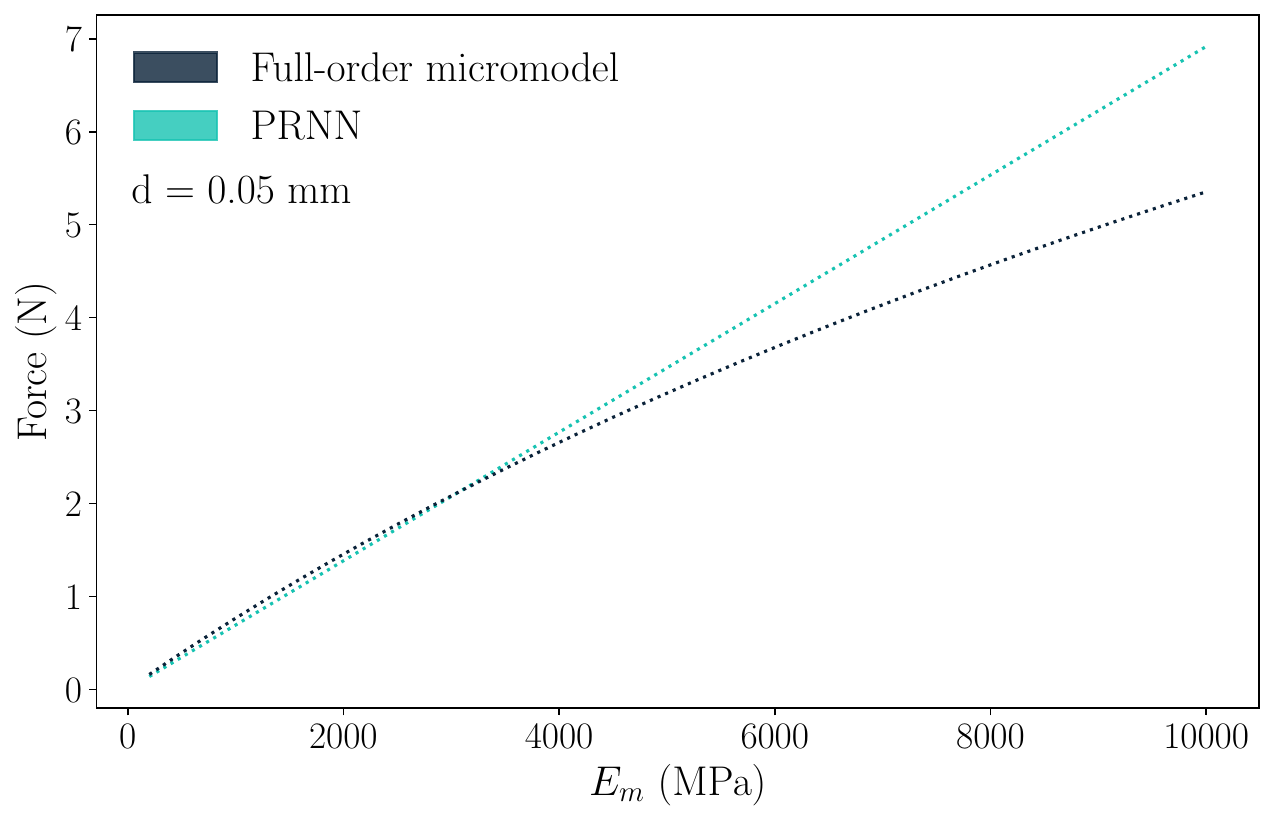}
        \caption{Forces obtained across the E$_m$ range at d = 0.05mm}
        \label{fig:emcomp}
\end{figure}


There is a notable deviation between the histograms resulting from the full-order and PRNN$_v$ driven models at lower displacement levels when $E_m$ is varied, shown in Figure \ref{fig:nnhfem1}. While PRNN$_v$ predicts a uniform force distribution for uniform $E_m$ input, the full-order UQ results indicate a nonlinear relation between $E_m$ and the linear elastic stiffness of the microstructure. This phenomenon is also reflected in Figure \ref{fig:emcomp}, where force is plotted for each $E_m$ value for both methods at the first time step: PRNN$_v$ shows a liner relation between $E_m$ and force at $d=0.05$ mm, while this relation is nonlinear for the full-order model. 

\begin{figure}[!h]
\centering
\includegraphics[scale=0.35]{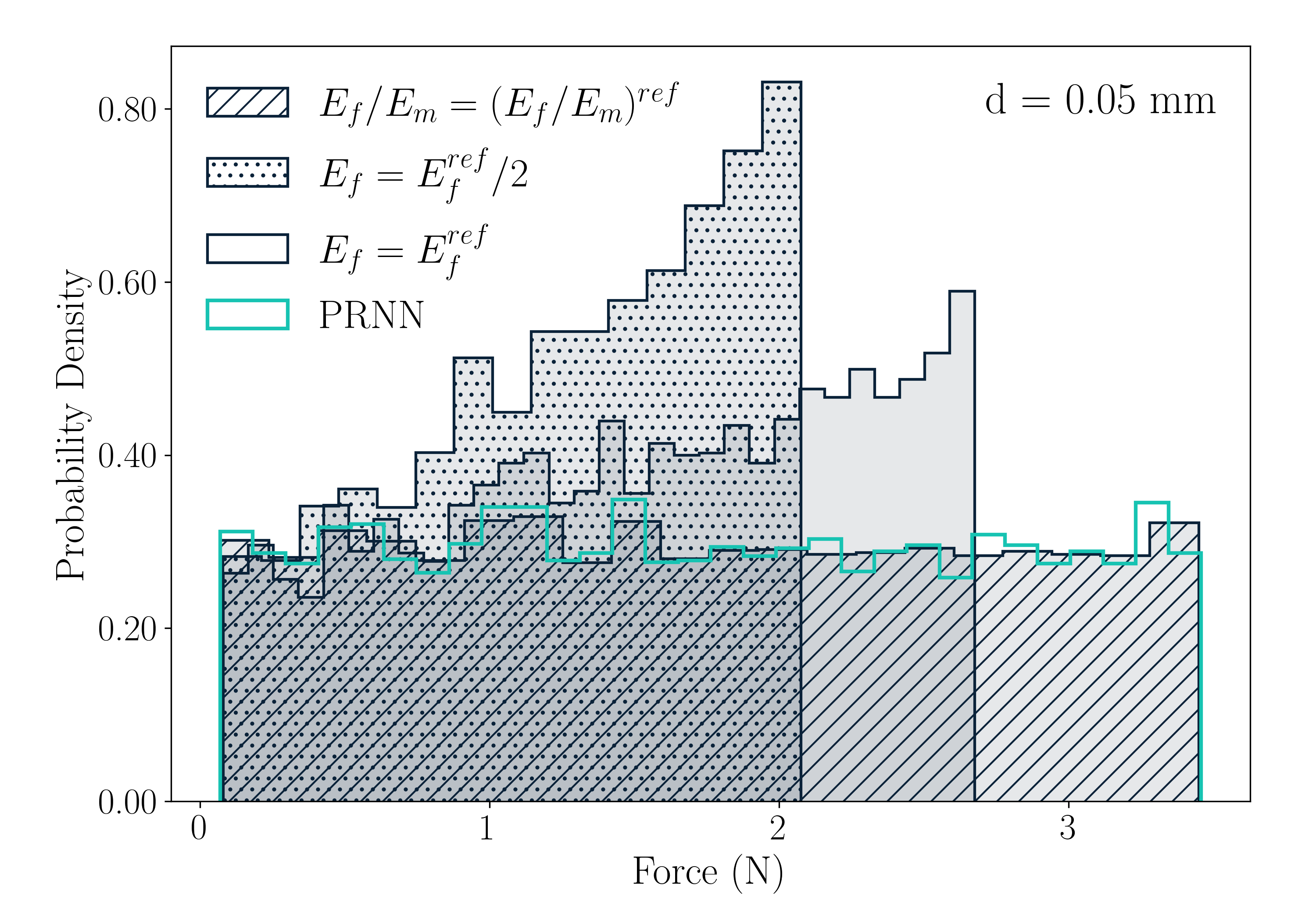}
\caption{\centering Force distribution from the PRNN-driven model and the full-order model with different E$_f$/E$_m$ ratios, at the first time step}
\label{fig:t1}
\end{figure}

This deviation can be explained by the matrix-fiber interaction at the microscale, which influences the homogenized elastic stiffness of the microstructure. This effect becomes more evident when varying the $E_f/E_m$ ratio in the full-order model. When $E_f$ is significantly larger than $E_m$, the variation in $E_m$ has a less significant effect on the stiffness of the microstructure. However, when $E_f$ is decreased to half of its reference value ($E_f^{ref} = 74000$ MPa), the increase in $E_m/E_f$ ratio results in a stronger influence of the matrix stiffness. This propagates through the multiscale framework, causing a more pronounced nonlinear distribution at the first time step, shown in Figure \ref{fig:t1}. When the $E_f/E_m$ ratio is kept constant at its reference value by also changing $E_f$ in the micromodel, the relationship between $E_m$ and the homogenized stiffness becomes linear. The resulting force distribution is shown in Figure \ref{fig:t1}, which almost perfectly aligns with the prediction of the previously presented PRNN$_v$-driven model. In absence of fiber material points in the used PRNN, the surrogate has no way to differentiate between these different scenarios and it defaults to assuming a linear dependence on $E_m$.



Although there is a small error in stiffness predictions which is reflected in the histograms at low displacement levels due to matrix-fiber interaction, its impact on the overall performance of the PRNN-driven UQ is low. Additionally, it is important to remember that the parameter intervals considered in this study are large. Overall, the predictive accuracy of the PRNN$_v$-driven model is excellent, as it is able to capture the highly nonlinear evolution in macroscopic force distributions.

\subsection{PRNN-driven uncertainty quantification on a fine mesh}

\begin{figure}[!h]
\centering
\begin{subfigure}[b]{1.\textwidth}
\centering
\includegraphics[scale=0.18]{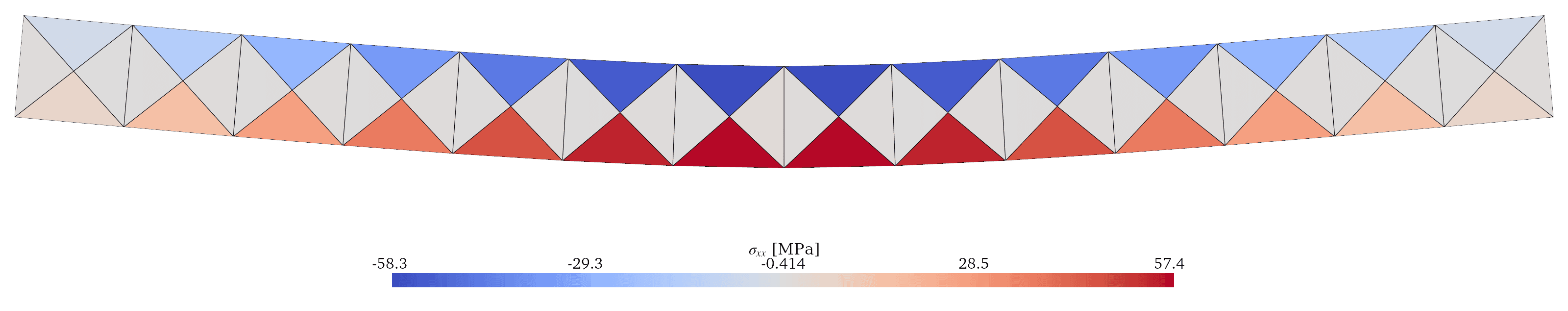}
\caption{\centering Coarse mesh solution}
\label{fig:tpbx}
\end{subfigure}
\vskip\baselineskip
\begin{subfigure}[b]{1.\textwidth}
\centering
\includegraphics[scale=0.18]{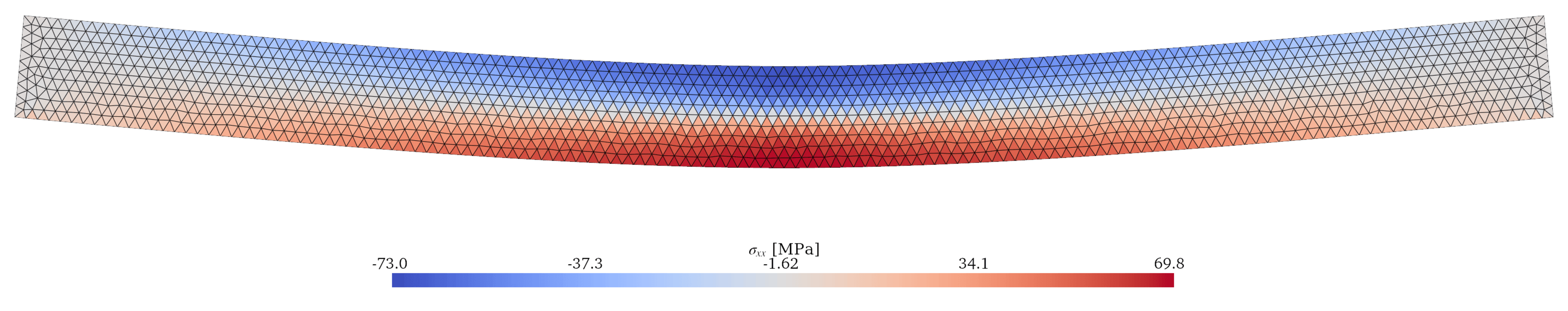}
\caption{\centering Fine mesh solution}
\label{fig:tpbfx}
\end{subfigure}
\caption{$\sigma_{xx}$ distributions over different mesh sizes for the full-order solution}
\label{fig:tpbxx}
\end{figure}

The uncertainty quantification performed on the coarse mesh demonstrates the network's ability to replace the full-order micromodel in a multiscale setting with varying microscale properties. Note that while this mesh is used for model comparison, it is too coarse to obtain accurate results. Figure \ref{fig:tpbxx} shows the $\sigma_{xx}$ distribution at the last time step for the full-order solution using the coarse and a more accurate finer mesh with 2690 elements, which highlights the need for using the finer mesh. However, obtaining the full-order results with this fine mesh can take up to 90 hours per simulation leading to a total of 51 years to run the complete Monte Carlo study, limiting its use for uncertainty quantification. On the other hand, the PRNN$_v$-driven model keeps the simulation time limited to approximately one minute, allowing for running a large number of simulations. This significant reduction in computational time makes it possible to perform UQ on the fine mesh. 


\begin{figure}[!h]
    \centering
    \begin{minipage}{\textwidth}
    \centering
    \begin{subfigure}{0.325\textwidth}
        \centering
        \includegraphics[width=\linewidth]{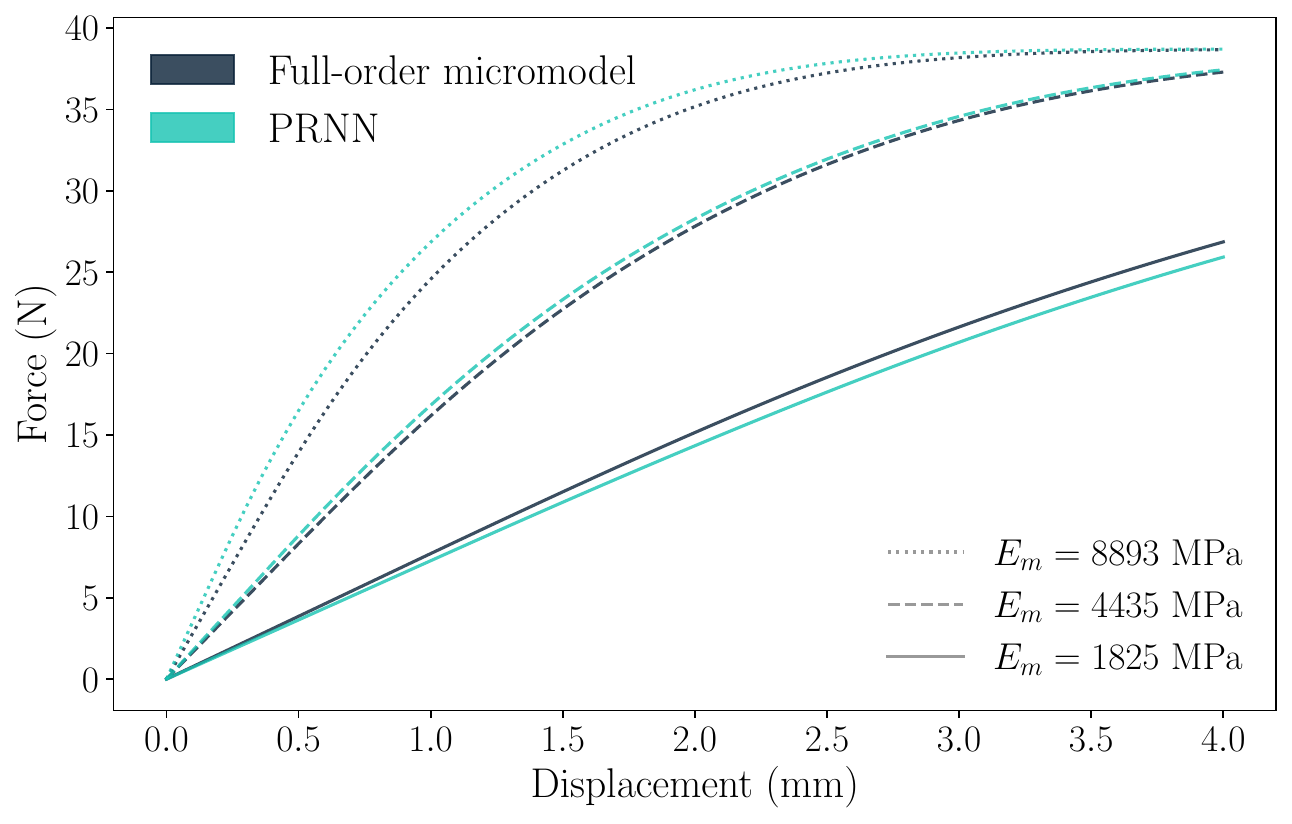}
        \caption{Varying E$_m$}
        \label{fig:emfc}
    \end{subfigure}
    \hfill
    \begin{subfigure}{0.325\textwidth}
        \centering
        \includegraphics[width=\linewidth]{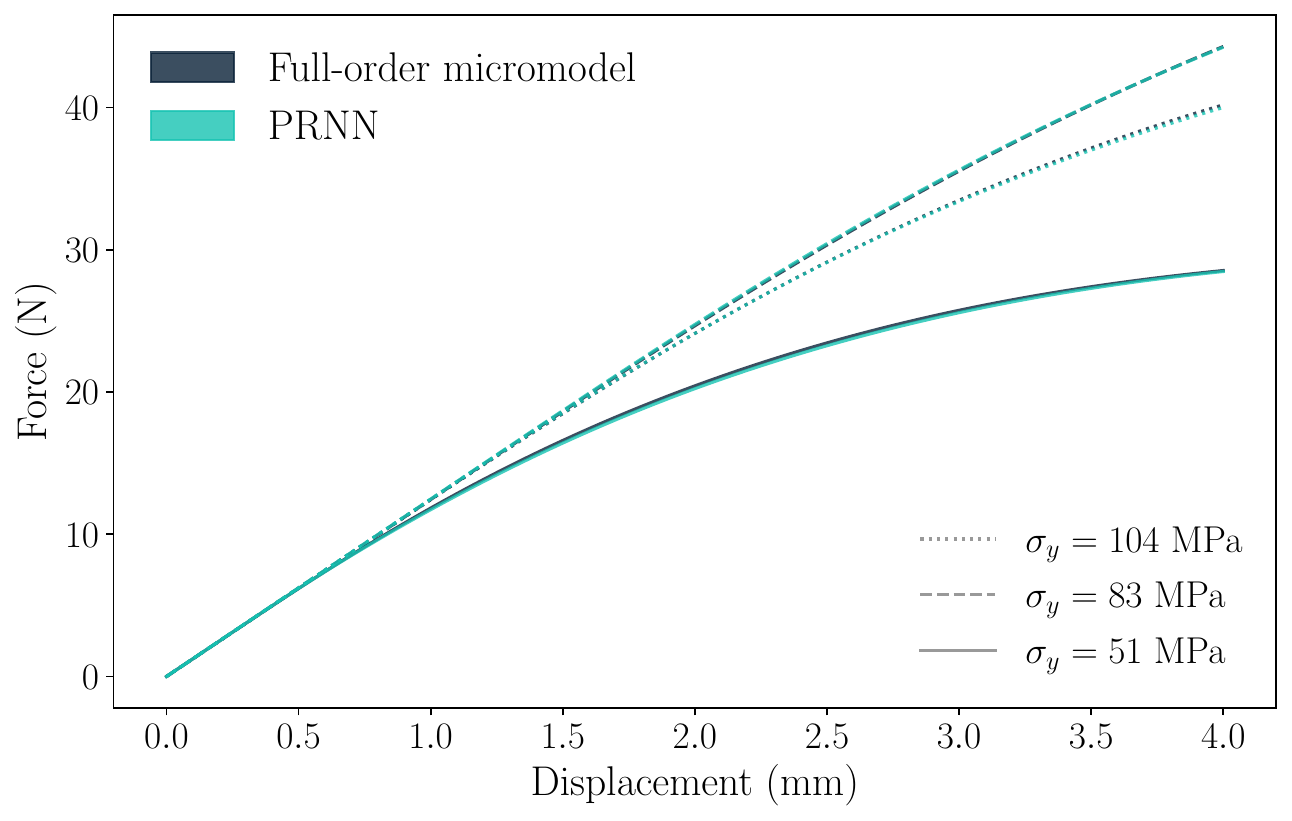}
        \caption{Varying $\sigma_y$}
        \label{fig:plcfc}
    \end{subfigure}
    \hfill
    \begin{subfigure}{0.325\textwidth}
        \centering
        \includegraphics[width=\linewidth]{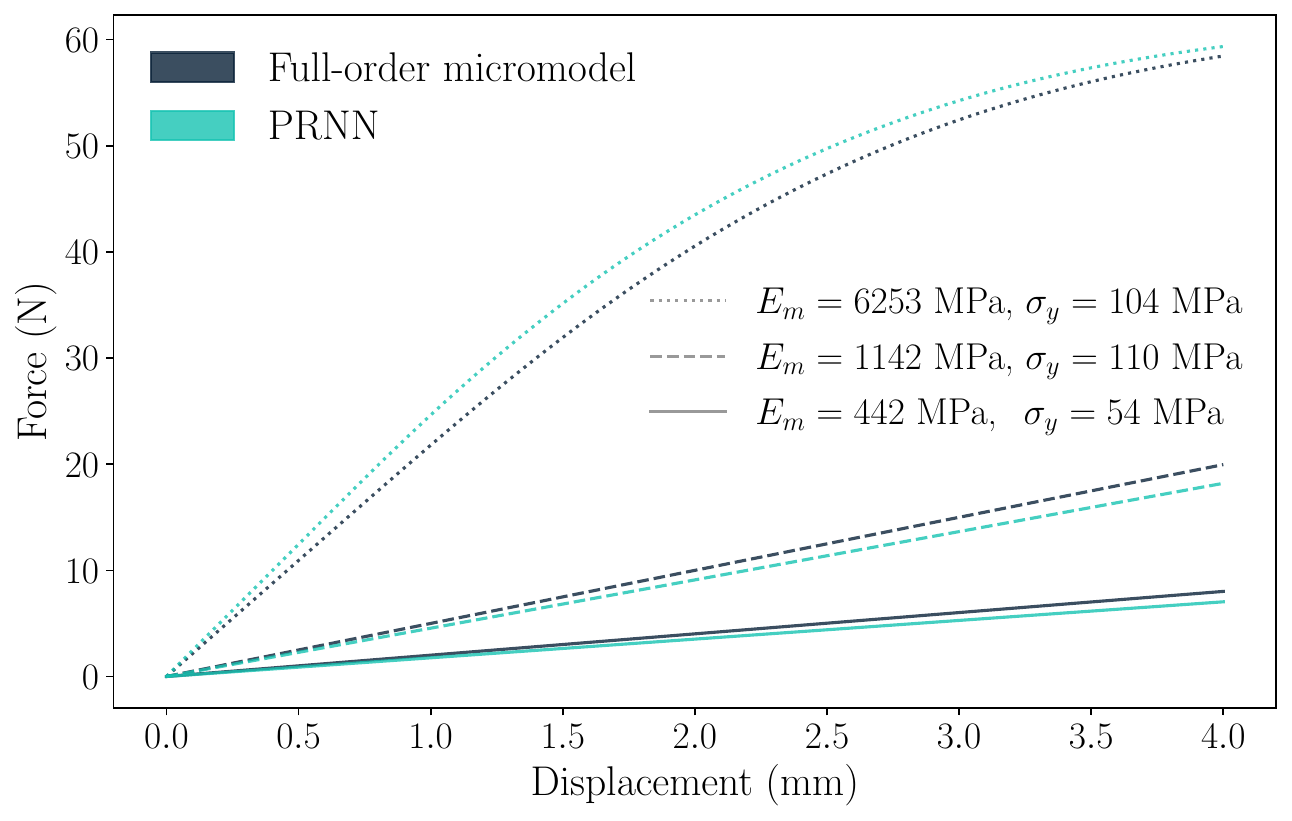} 
        \caption{Varying E$_m$ and $\sigma_y$ simultaneously}
        \label{fig:emplcfc}
    \end{subfigure}
    \end{minipage}
    \caption{\centering Comparing the load-displacement curve resulting from various parameters using the full-order and the PRNN-driven model}
    \label{fig:fc}
\end{figure}

Before performing UQ on the finer mesh with the PRNN$_v$-driven model only, a brief validation is conducted to evaluate whether the accuracy observed previously on the coarse mesh is maintained. For this, 3 values are selected from each parameter set and full order analyses are ran using these sets. These results are then compared to the corresponding PRNN$_v$-driven predictions, shown in Figure \ref{fig:fc}. The close agreement between the two methods confirms the accuracy across different mesh sizes, confirming its reliability to perform a UQ on the fine mesh.

\begin{figure}[!h]
  \centering
  \begin{subfigure}[b]{0.32\textwidth}
    \centering
    \includegraphics[width=\textwidth]{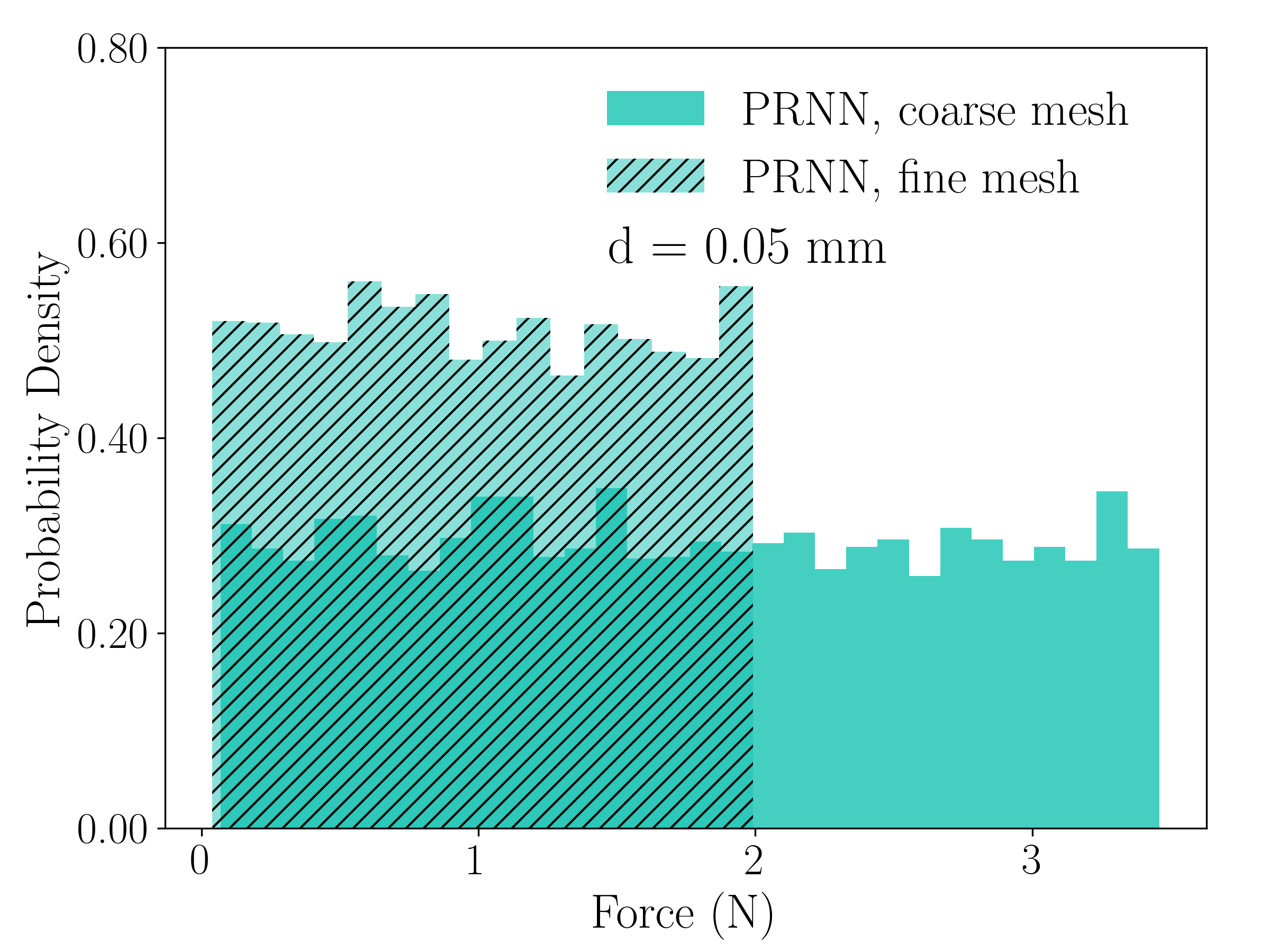}
    \caption{\centering Displacement = 0.05 mm}
    \label{fig:nnem1}
  \end{subfigure}
  \hfill
  \begin{subfigure}[b]{0.32\textwidth}
    \centering
    \includegraphics[width=\textwidth]{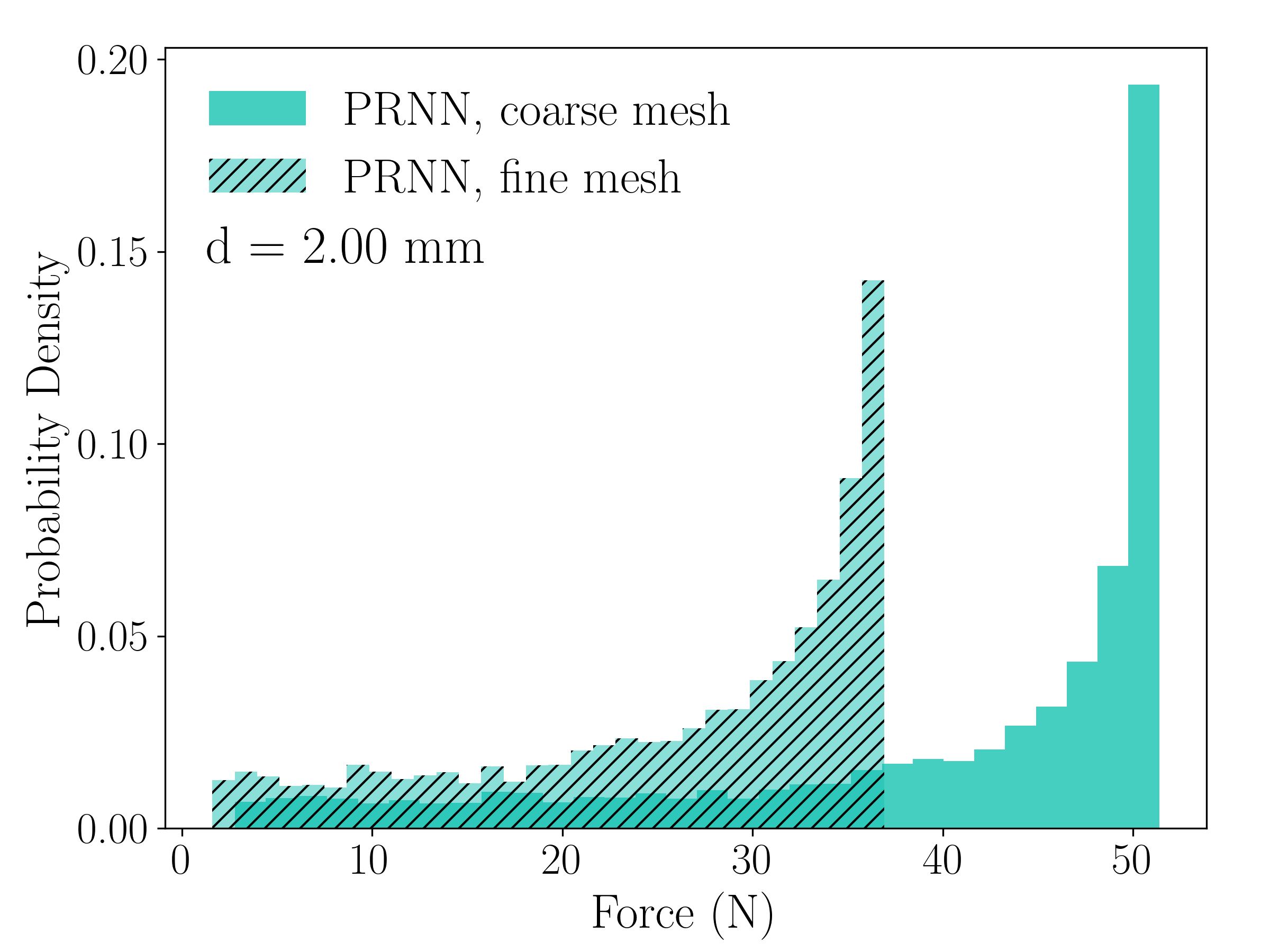}
    \caption{\centering Displacement = 2 mm}
    \label{fig:nnem2}
  \end{subfigure}
  \hfill
  \begin{subfigure}[b]{0.32\textwidth}
    \centering
    \includegraphics[width=\textwidth]{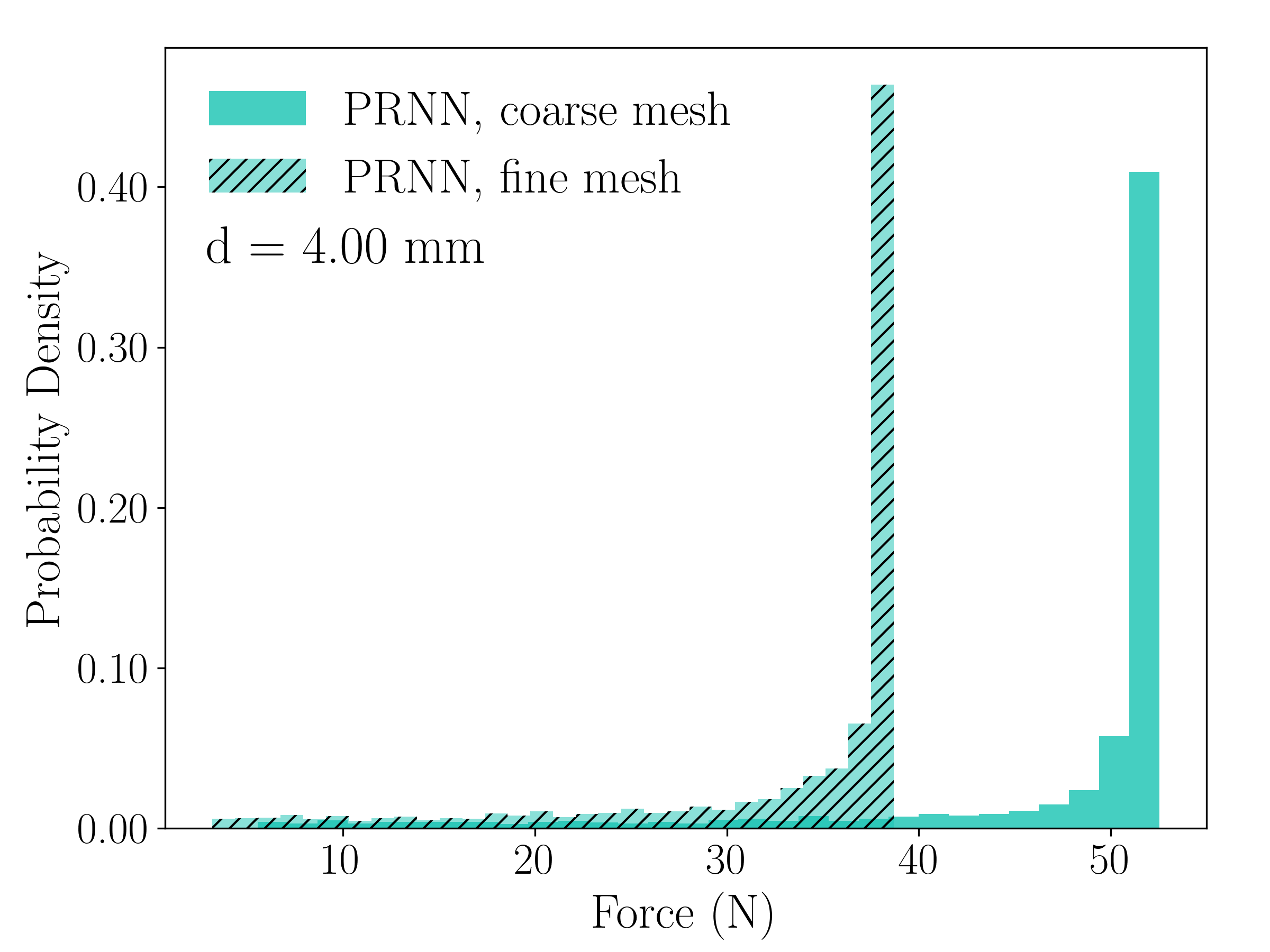}
    \caption{\centering Displacement = 4 mm}
    \label{fig:nnem3}
  \end{subfigure}
\caption{\centering Load distributions at the midpoint of the beam for varying E$_m$, using a fine mesh}
  \label{fig:nnem}
\end{figure}

\begin{figure}[!h]
  \centering
  \begin{subfigure}[b]{0.32\textwidth}
    \centering
    \includegraphics[width=\textwidth]{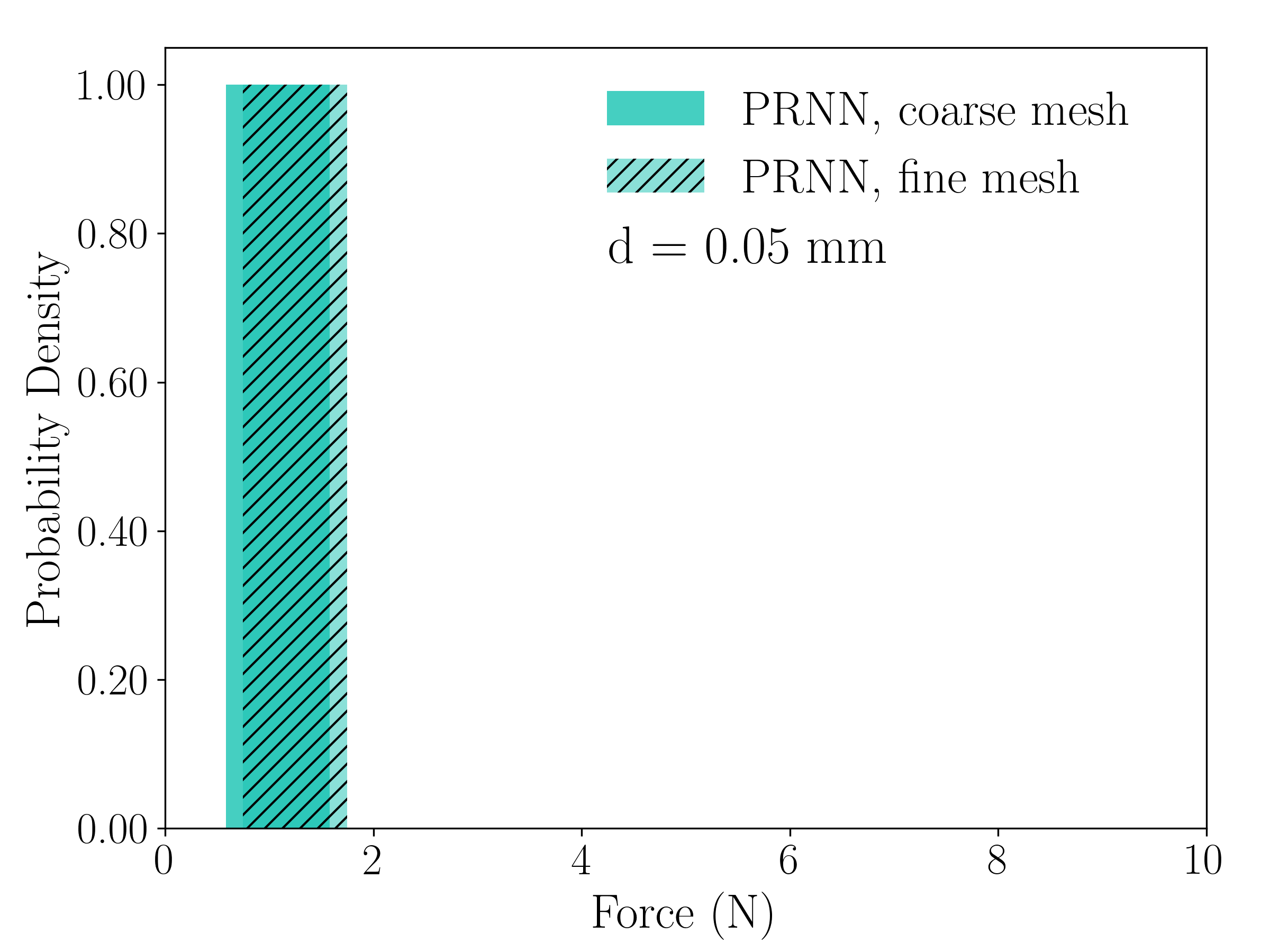}
    \caption{\centering Displacement = 0.05 mm}
    \label{fig:nnplc1}
  \end{subfigure}
  \hfill
  \begin{subfigure}[b]{0.32\textwidth}
    \centering
    \includegraphics[width=\textwidth]{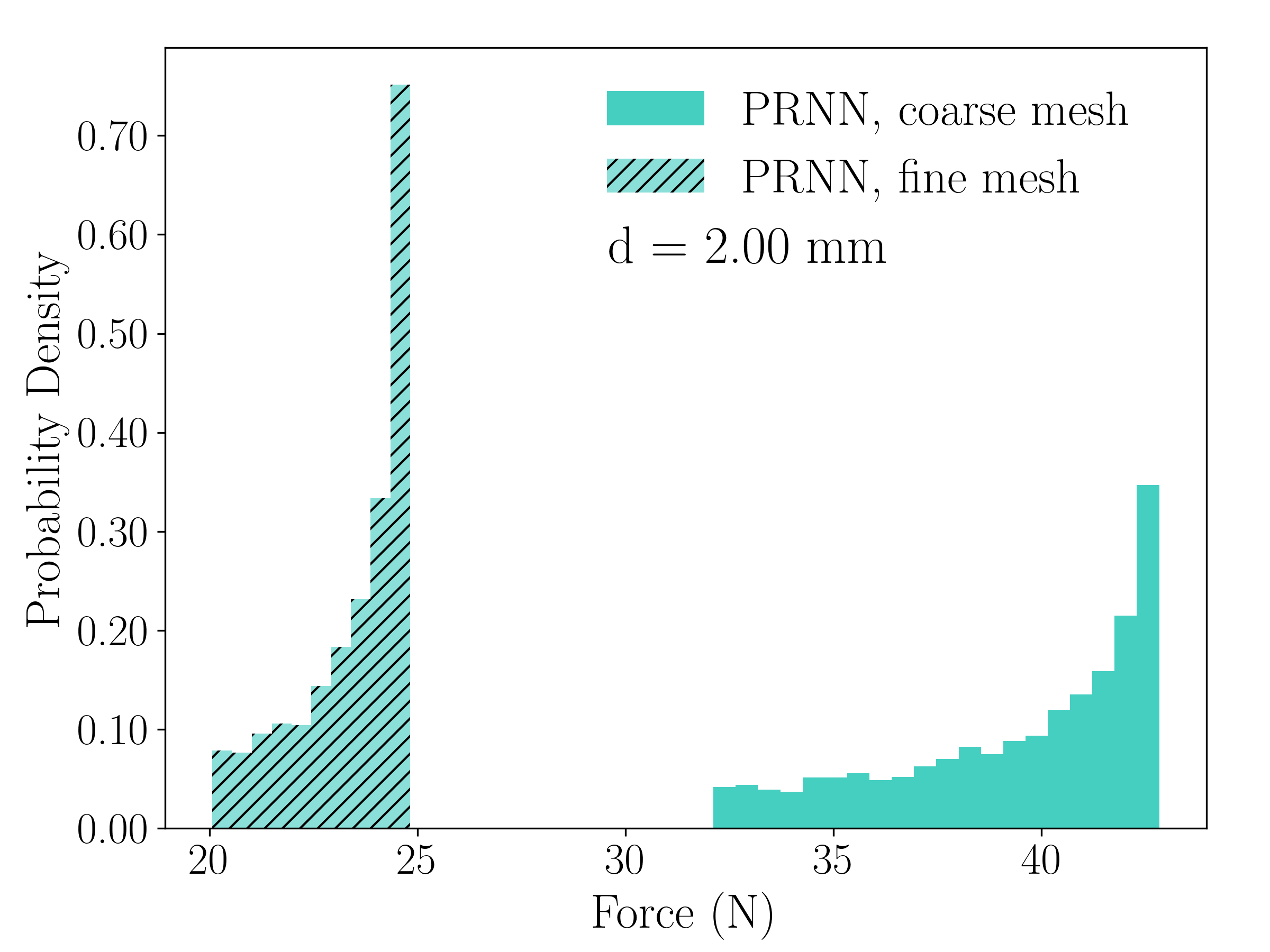}
    \caption{\centering Displacement = 2 mm}
    \label{fig:nnplc2}
  \end{subfigure}
  \hfill
  \begin{subfigure}[b]{0.32\textwidth}
    \centering
    \includegraphics[width=\textwidth]{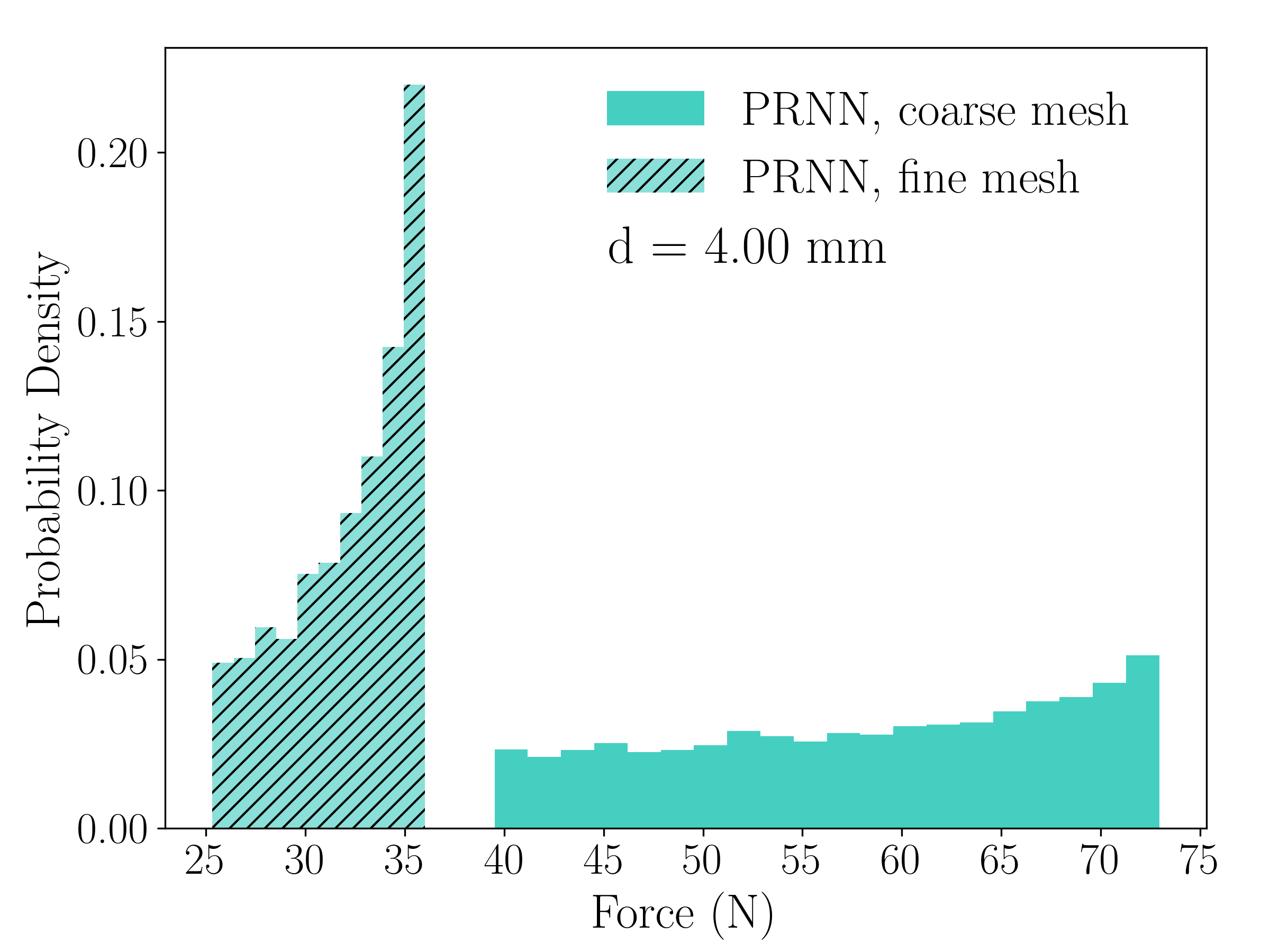}
    \caption{\centering Displacement = 4 mm}
    \label{fig:nnplc3}
  \end{subfigure}
\caption{\centering Load distributions at the midpoint of the beam for varying $\sigma_y$, using a fine mesh}
  \label{fig:nnplc}
\end{figure}

\begin{figure}[!h]
  \centering
  \begin{subfigure}[b]{0.32\textwidth}
    \centering
    \includegraphics[width=\textwidth]{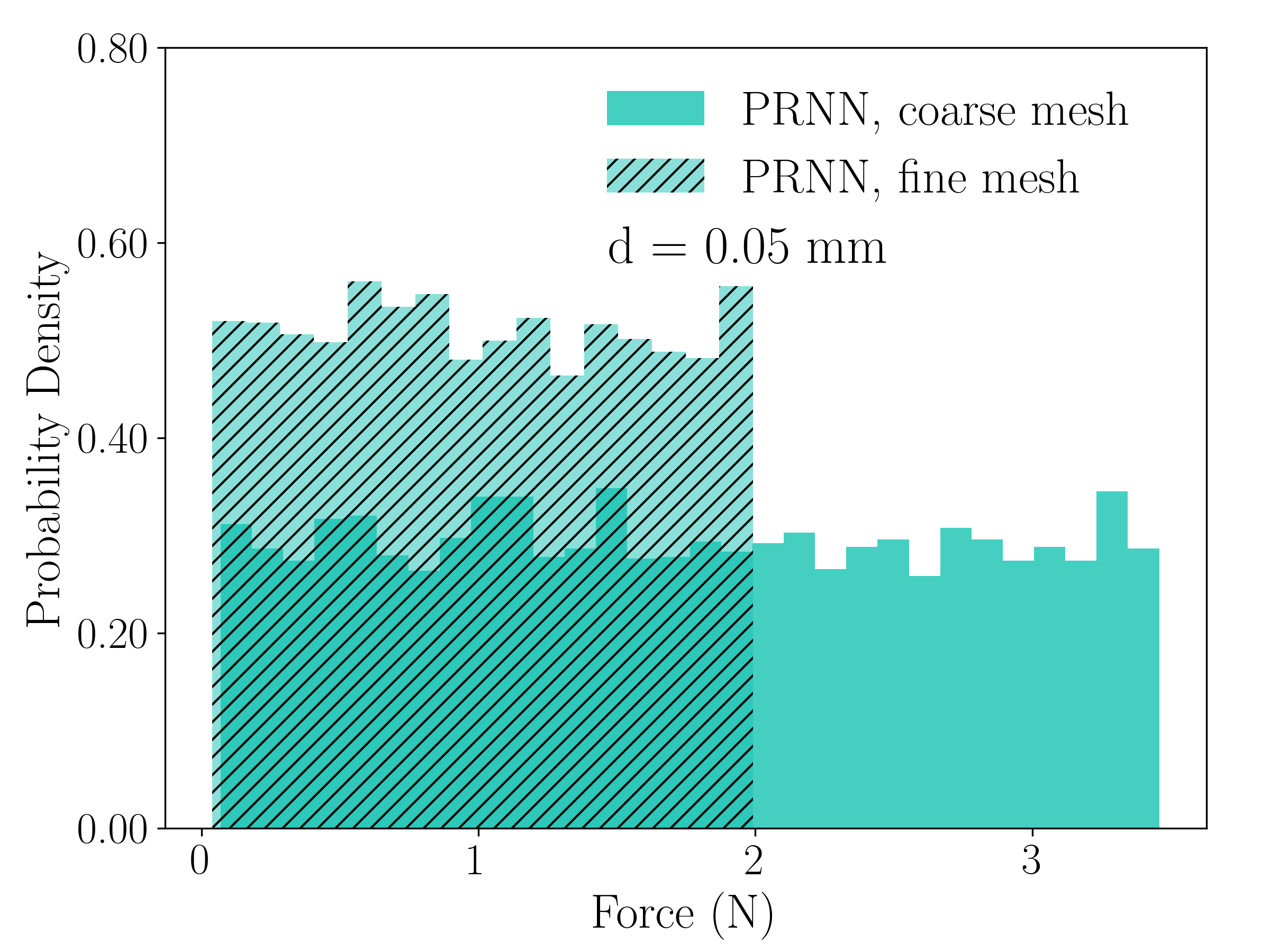}
    \caption{\centering Displacement = 0.05 mm}
    \label{fig:nnemplc1}
  \end{subfigure}
  \hfill
  \begin{subfigure}[b]{0.32\textwidth}
    \centering
    \includegraphics[width=\textwidth]{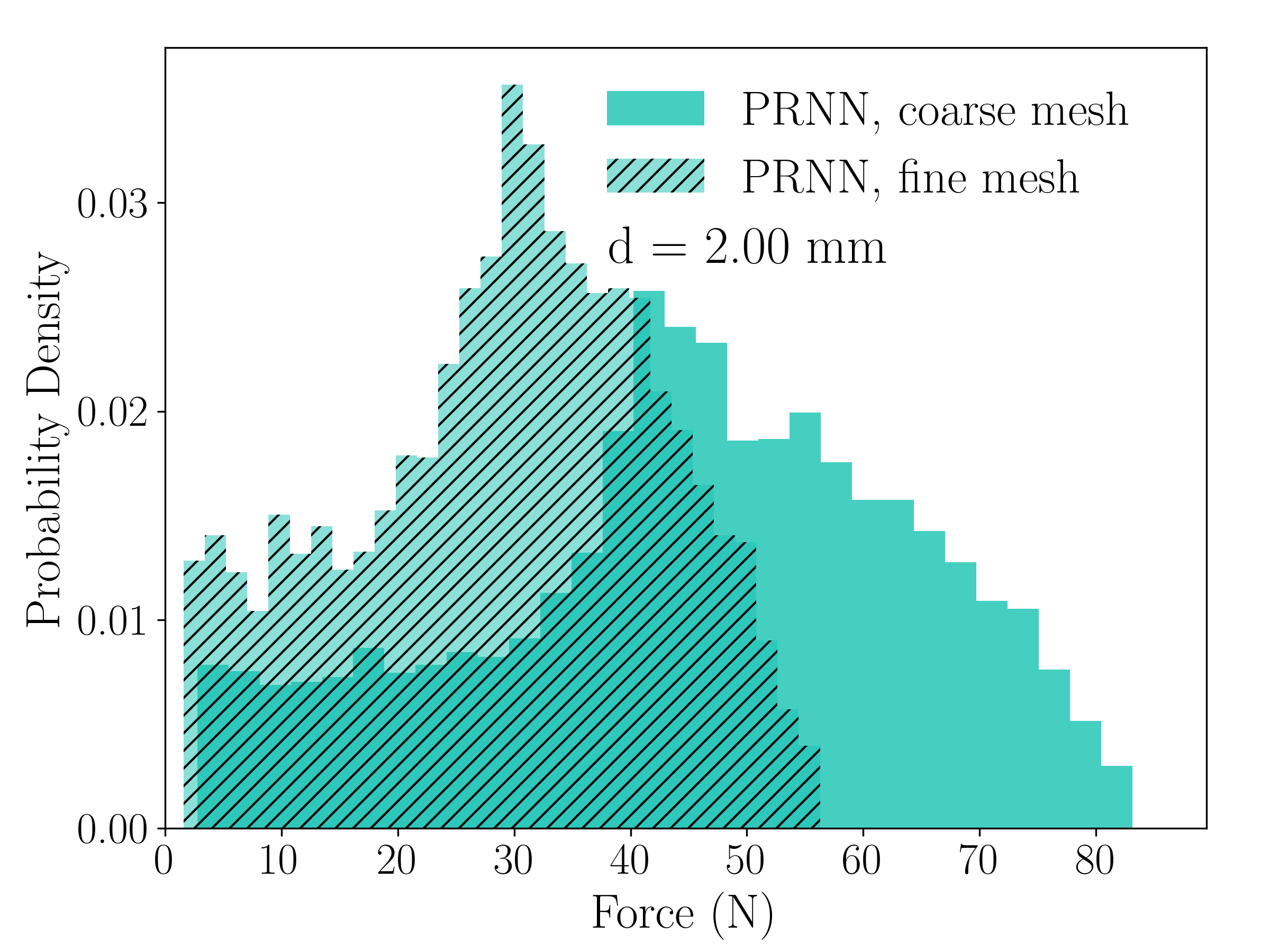}
    \caption{\centering Displacement = 2 mm}
    \label{fig:nnemplc2}
  \end{subfigure}
  \hfill
  \begin{subfigure}[b]{0.32\textwidth}
    \centering
    \includegraphics[width=\textwidth]{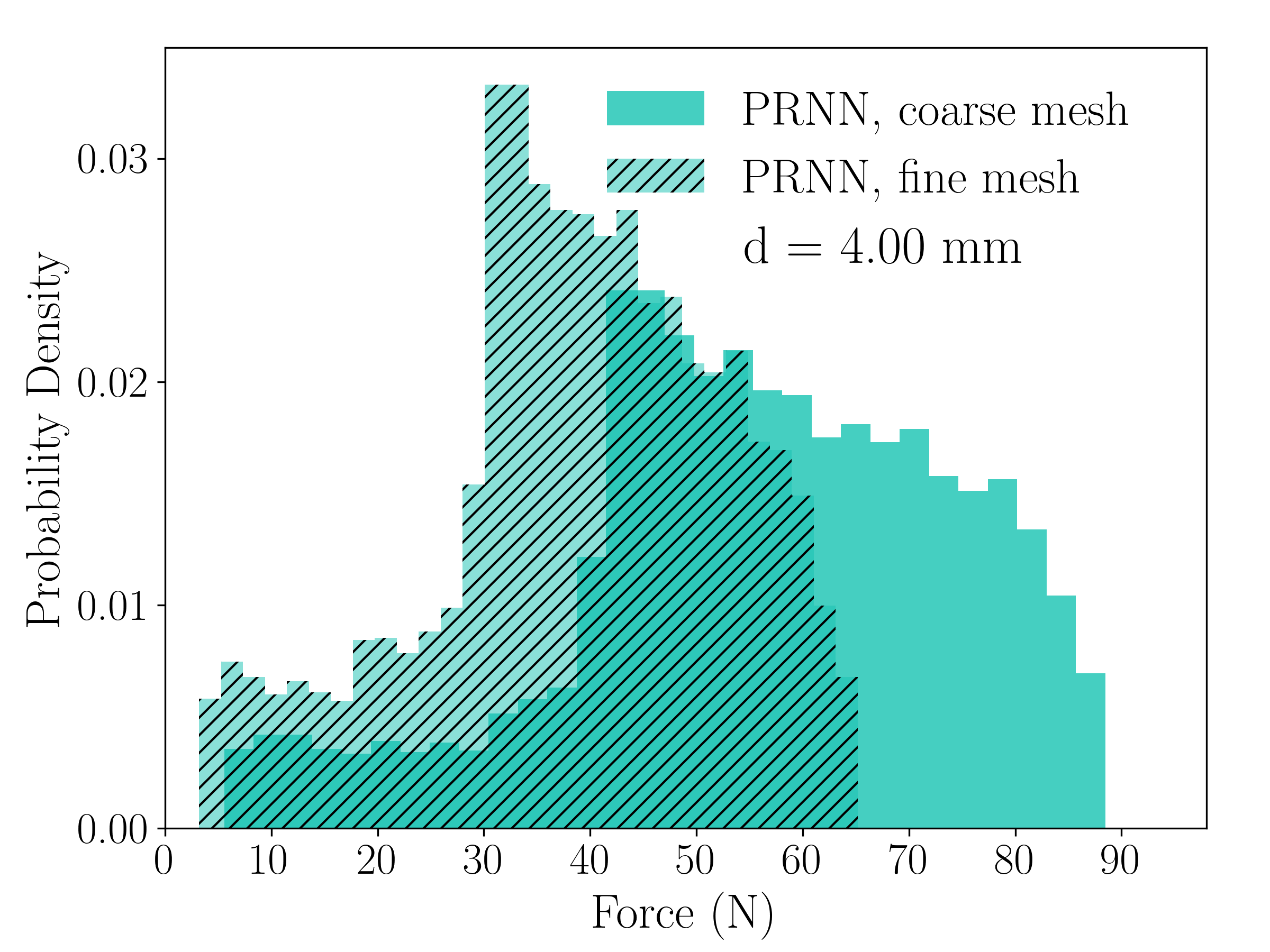}
    \caption{\centering Displacement = 4 mm}
    \label{fig:nnemplc3}
  \end{subfigure}
\caption{\centering Load distributions at the midpoint of the beam for varying E$_m$ and $\sigma_y$ simultaneously, using a fine mesh}
  \label{fig:nnemplc}
\end{figure}

Finally, uncertainty quantification is performed for the same 5000 samples drawn from a uniform distribution as earlier. Force distribution histograms for varying $E_m$, $\sigma_y$, or both at the same time are shown in Figures \ref{fig:nnem}-\ref{fig:nnemplc} at three time steps, comparing the results to the ones obtained from the PRNN$_v$ driven model on the coarse mesh. Note that the shape of the distributions obtained on the coarse mesh are similar to those obtained from the fine one, however, the coarse mesh simulations consistently overpredict the forces, further highlighting the need for using the fine mesh for accurate results.

\newpage
~\newpage
\section{Conclusions}
\label{}
In this paper, we investigated whether Physically Recurrent Neural Networks trained on responses obtained from a fixed RVE can accurately extrapolate to varying microscale material properties without retraining, and if it can be used in multiscale uncertainty quantification. The PRNN was previously shown to accurately replace the full-order RVE in a multiscale setting for a fixed set of parameters \cite{MAIA2023115934}. Due to the presence of material models in their material layer, PRNNs offer the possibility to modify these parameters after training, without modifying the weights or the architecture of the network. 

To obtain the network with fixed architecture and weights, the PRNN was trained with data generated from a micromodel with fixed parameters. A sensitivity analysis revealed the Young's modulus and yield strength of the matrix to have largest influence on the homogenized microscale response. Therefore, micromodels with a variety of these parameters were generated to obtain the dataset used for testing the network. A wide range was used for the parameters in order to explore the limits of the network.

First, the prediction of the network at the microscale was evaluated. Two ways of testing the PRNN were considered: the parameters in the material layer were either kept fixed (PRNN$_f$) or they were dynamically adjusted according to the input data (PRNN$_v$). As expected, PRNN$_f$ predicted poorly for a large variation in material properties. On the other hand, PRNN$_v$ showed high accuracy across the board, whether varying one parameter at a time or both simultaneously. This showed its ability to generalize to variations in material parameters without retraining the network.

Motivated by the exceptional performance at the microscale, multiscale UQ was considered next. A three-point bending case was simulated using a coarse mesh so that a comparison could be made between the full-order multiscale and PRNN-driven simulations. The PRNN-driven model was able to capture highly nonlinear force distributions and their transition over time. A difference in homogenized stiffness prediction was observed that arises from the nonlinear fiber-matrix interaction. A brief analysis on the effect of the E$_m$/E$_f$ proportion on the homogenized stiffness revealed that if this proportion is kept constant to the one used during training, this error diminishes. The overall UQ performance of the PRNN-driven model is however not significantly affected by this error. The nonlinear evolution of the distribution in macroscopic load required to cause increasing displacement levels is accurately captured across a wide range of displacement levels and parameter values.

Finally, a PRNN-driven UQ was conduced on a much finer mesh that would otherwise be infeasable with the full-order model due to the computational time it is associated with (90 hours per simulation). Using PRNN$_v$ at the integration points decreased this run time to about one minute per simulation, allowing for an extensive UQ with a more accurate mesh size. This marks a milestone in applying a physics-based surrogate in a full-scale Monte Carlo multiscale UQ setting, demonstrating feasibility with high accuracy. The PRNN-driven predictions were compared with the results of the full-order model for a small amount of selected parameters to verify the performance on the fine mesh. These results show the ability of the PRNN as a surrogate at UQ in multiscale modeling of composites, reducing the computational costs to the extent that what was once impossible now can be done at a reasonable cost, with excellent accuracy.



\bibliographystyle{elsarticle-num}  
\bibliography{report}


\end{document}